\documentclass[%
 reprint,
 amsmath,amssymb,
 aps,
]{revtex4-1}

\usepackage{graphicx}
\usepackage{latexsym}
\usepackage{graphics}
\usepackage{textcomp}
\usepackage{amsmath}
\usepackage{amssymb}

\graphicspath{{figures/}}
\usepackage{xspace}
\newcommand{\um}{\,\ensuremath{\mu\text{m}}\xspace}
\newcommand{\percent}{\,\%\xspace}

\begin{document}
% =============== Title, authors, addresses ===============
\title{Optical nanofibers and spectroscopy}

\author{R. Garcia-Fernandez}
\affiliation{Institut f\"{u}r Physik, Johannes Gutenberg-Universit\"{a}t Mainz, 55099 Mainz, Germany}
\affiliation{Vienna Center for Quantum Science and Technology, Atominstitut, TU Wien, 1020 Wien, Austria} 
\author{W. Alt}
\affiliation{Institut f\"{u}r Angewandte Physik, Universit\"{a}t Bonn, Wegelerstr. 8, 53115 Bonn, Germany}
\author{F. Bruse}
\affiliation{Institut f\"{u}r Angewandte Physik, Universit\"{a}t Bonn, Wegelerstr. 8, 53115 Bonn, Germany}
\author{C. Dan}
\affiliation{Institut f\"{u}r Angewandte Physik, Universit\"{a}t Bonn, Wegelerstr. 8, 53115 Bonn, Germany}
\author{K. Karapetyan}
\affiliation{Institut f\"{u}r Angewandte Physik, Universit\"{a}t Bonn, Wegelerstr. 8, 53115 Bonn, Germany}
\author{O. Rehband}
\affiliation{Institut f\"{u}r Physik, Johannes Gutenberg-Universit\"{a}t Mainz, 55099 Mainz, Germany}
\author{A. Stiebeiner}
\affiliation{Institut f\"{u}r Physik, Johannes Gutenberg-Universit\"{a}t Mainz, 55099 Mainz, Germany}
\affiliation{Vienna Center for Quantum Science and Technology, Atominstitut, TU Wien, 1020 Wien, Austria} 
\author{U.Wiedemann}
\affiliation{Institut f\"{u}r Angewandte Physik, Universit\"{a}t Bonn, Wegelerstr. 8, 53115 Bonn, Germany}
\author{D. Meschede}
\affiliation{Institut f\"{u}r Angewandte Physik, Universit\"{a}t Bonn, Wegelerstr. 8, 53115 Bonn, Germany}
\email{meschede@uni-bonn.de}
\author{A. Rauschenbeutel}
\affiliation{Institut f\"{u}r Physik, Johannes Gutenberg-Universit\"{a}t Mainz, 55099 Mainz, Germany}
\affiliation{Vienna Center for Quantum Science and Technology, Atominstitut, TU Wien, 1020 Wien, Austria} 
\email{arno.rauschenbeutel@ati.ac.at}
   
\date{\today}

% ======================== Abstract =============================
\begin{abstract}
We review our recent progress in the production and characterization of
tapered optical fibers with a sub-wavelength diameter waist. Such fibers
exhibit a pronounced evanescent field and are therefore a useful tool for
highly sensitive evanescent wave spectroscopy of adsorbates on the fiber
waist or of the medium surrounding. We use a carefully
designed flame pulling process that allows us to realize preset fiber
diameter profiles. In order to determine the waist diameter and to verify
the fiber profile, we employ scanning electron microscope measurements and a
novel accurate \textit{in situ} optical method based on harmonic generation. We use our fibers for
linear and non-linear absorption and fluorescence spectroscopy of
surface-adsorbed organic molecules and investigate their agglomeration
dynamics. Furthermore, we apply our spectroscopic method to quantum dots on
the surface of the fiber waist and to caesium vapor surrounding the fiber.
Finally, towards dispersive measurements, we present our first results on
building and testing a single-fiber bi-modal interferometer.

\end{abstract}

\maketitle

% ======================== 1. Introduction =============================
\section{Introduction}
\label{sec:intro}

Glass fibers are nowadays well established as the backbone of the modern communication society. In a standard glass fiber used for telecommunications, the light is guided inside the core and is isolated from the environment by the cladding, allowing low loss and low disturbance transmission over long distances \cite{snyder}. In contrast, optical nanofibers \cite{brambilla} with a diameter smaller than the wavelength of the guided light exhibit a strong lateral confinement of the guided mode in conjunction with a pronounced evanescent field surrounding the fiber \cite{le kien}. These properties make optical nanofibers a unique tool for efficient and controlled coupling of light with matter on or near their surface. A wide range of nanofiber applications are emerging in fields like, e.g., optical sensing \cite{villatoro,polynkin,zhang}, nanofiber-based evanescent wave spectroscopy \cite{absorption,fluorescence}, nonlinear optics \cite{birks0,grubsky,gattass,shg}, cold atom physics \cite{sague,nayak,nayak2,morrissey,vetsch}, probing of nano-cavities and cavity quantum electrodynamics (cQED) \cite{spillane,srinivasan,barclay,srinivasan2}, coupling to nano-cavity lasers \cite{min} or single quantum dot spectroscopy \cite{srinivasan3}. For these applications, high-quality nanofibers with a high diameter uniformity and low surface roughness are required, rendering the technical realization a demanding process. 

A standard fabrication method for optical nanofibers is flame pulling of commercial single mode optical fibers resulting in tapered optical fibers (TOFs) with a nanofiber waist \cite{tong,brambilla2,clohessy,sumetsky,warken,pricking}. Efficient coupling of the light into and out of the nanofiber waist requires a careful design of the taper profile \cite{love,transmission}. Therefore, we first describe the production of nanofibers with our computer controlled fiber pulling rig as well as their main properties in Sect.~\ref{sec:fibers}. In order to verify the accuracy and precision of our production process, we present scanning electron microscope measurements of the TOF profile and compare the experimental results with the theoretically predicted target profile. Further, in Sect.~\ref{sec:characterization} we present a method for measuring the diameter of the nanofiber waist, which is based on second and third harmonic generation and has an accuracy of better than 2\percent \cite{shg}. 

In Sect.~\ref{sec:surface} of this paper, we focus on applications of nanofibers for surface spectroscopy. Fiber-based evanescent wave spectroscopy (EWS) offers stability, fast response and applications with \textit{in situ} measurements. Motivated by these advantages, fiber-based EWS sensors have been demonstrated \cite{potyrailo}. It is remarkable that the sensitivity of nanofiber spectroscopy depends strongly on the fiber diameter $d$ and the fiber length $L$: At resonance, the absorption probability per surface adsorbed particle with absorption cross section $\sigma$ can be roughly estimated to be $\sigma /\lambda^2$ because, in the fiber waist of a silica nanofiber, the light field can be confined to an effective cross section on the order of $\lambda^2$, similar to a strongly focused light beam. In contrast to the situation in free space, this confinement is however maintained along the entire fiber length $L$, and hence the typical depth of focus of a free space beam is enhanced by 3 to 5 orders of magnitude. As we show in detail in Sect.~\ref{theory}, the spectroscopic sensitivity $\eta$ of the nanofiber scales as 

\begin{equation} 
\eta(\lambda) \approx \eta_{\rm free}(\lambda) \frac{ \pi d L}  {\lambda^2} = \eta_{\rm free}(\lambda) \frac{ \pi L}{\lambda}, 
\end{equation}
where we assumed $d = \lambda$ and $\eta_{\rm free}$ is the value for a free space beam. Therefore, the strongly confined light in the nanofiber can be considered as an extremely elongated ``infinite'' focus. We demonstrate the versatility of fiber-based EWS by a series of absorption and fluorescence measurements performed on systems like surface-adsorbed organic molecules or quantum dots. Further, fiber-based EWS is also suited for spectroscopic measurements of atoms surrounding the nanofiber, as shown in Sect.~\ref{sec:cs} on the example of caesium vapor.

Finally, we use nanofibers for the realization of an interferometric sensor based on the interference of two co-propagating transversal modes along the TOF \cite{bures,orucevic}. The first experimental results are presented in Sect.~\ref{sec:interferometer}. This technique is highly sensitive to changes in the refractive index in the vicinity of the nanofiber surface and therefore opens the route towards dispersive measurements on surface adsorbates \cite{lou}.

% ====================== 2. Manufacturing, simulation and properties ==========================
\section{Fabrication and design of nanofibers}
\label{sec:fibers}

In the following, we summarize the fabrication procedure of our nanofibers. 

% ====================== 2.1 Manufacturing ==========================
\subsection{Description of the fiber pulling rig}
\label{pulling}

\begin{figure}[h]
\centering
  \resizebox{0.48\textwidth}{!}{%
  \includegraphics{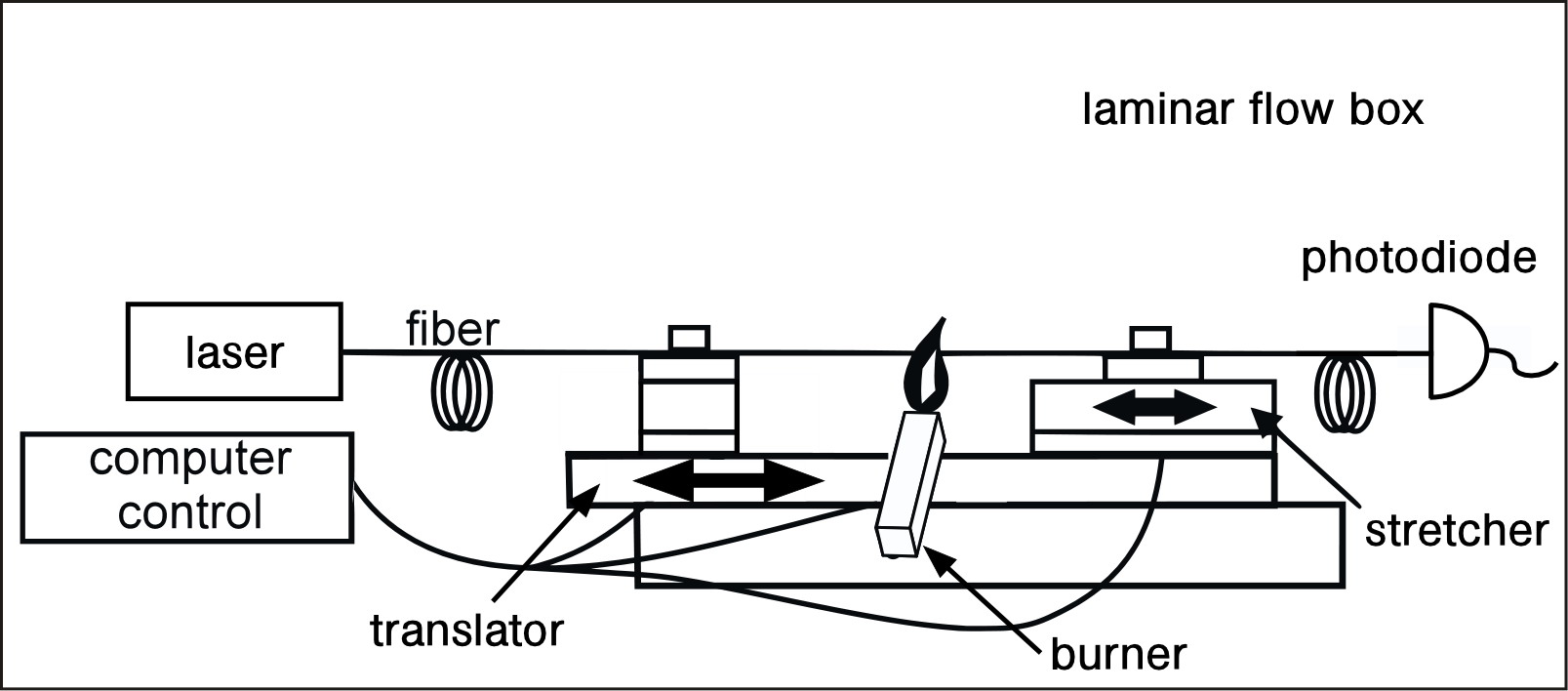}
}
    \caption{Schematic of the fiber pulling rig. A standard glass fiber is clamped
on two stacked translation stages, which move and stretch the glass fiber
which is heated with the flame of a stationary gas burner. The pulling procedure is performed
under clean room conditions in a laminar flow box.}\label{Fig_pulling}
\end{figure}

All our experiments are performed using the nanofiber waist of tapered optical fibers (TOFs). We produce the TOFs by stretching a standard optical fiber while heating it with a traveling hydrogen/oxygen flame \cite{birks}. A schematic of the fiber pulling rig is shown in Fig.~\ref{Fig_pulling}. It consists of two linear translation stages, the translator and the stretcher, which allow sub-micrometer positioning, and is equipped with precisely parallel aligned V-groove holders to which the fiber is clamped and tightened. A computer-controlled hydrogen/oxygen burner, located between these two holders, provides a clean flame. The flame heats a section of the fiber with a length of approximately 1~mm to a temperature of about $	1500\,^{\circ}\mathrm{C}$, making it soft and malleable. The translator moves the fiber back and forth above the flame while the stretcher simultaneously elongates it. Due to volume conservation, the fiber diameter is thus gradually reduced. We use an analytic algorithm in order to calculate the trajectories of the translator and of the stretcher to obtain a predetermined fiber shape. This process allows us to fabricate TOFs with a homogeneous waist of diameters down to 100~nm and a typical length of 1--10~mm \cite{warken}. The transmission of light through the fiber is continuously monitored during the pulling process, yielding valuable information for quality control of the taper and its overall transmission. For monochromatic light of 850~nm wavelength and a nanofiber with a 500-nm diameter waist, our fabrication procedure yields up to $98.7\%$ of the initial fiber transmission.

% ====================== 2.2 Simulation and properties ==========================
% bonn_simulation_and_properties
\subsection{Properties of the guided light}

\begin{figure}[h]
\centering
  \resizebox{0.48\textwidth}{!}{%
  \includegraphics{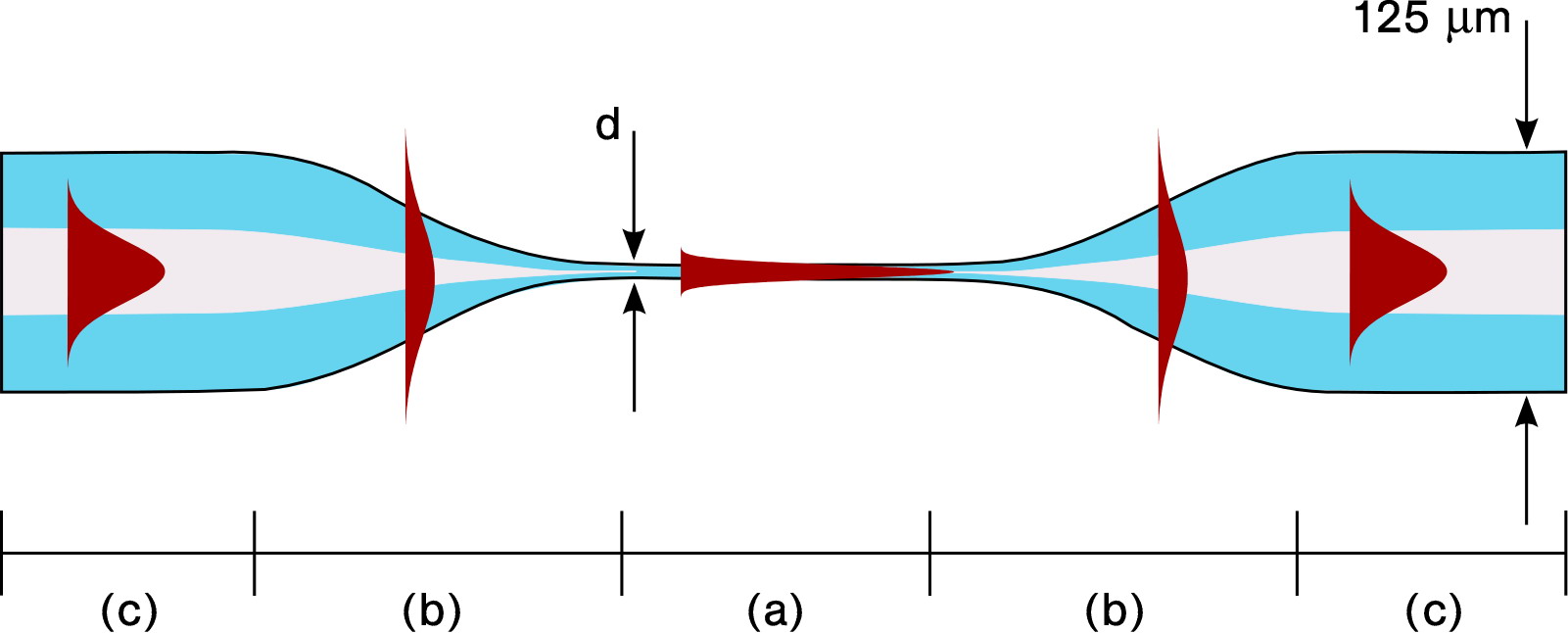}
}
    \caption{Schematic of a tapered optical fiber (TOF) showing the light propagation
and conversion of the fundamental mode (c) in the tapered region (b) to the
nanofiber waist (a) and back again to the initial mode (c). The intensity
profile of the guided mode is schematically represented by the filled curves.}\label{Fig_fiber}
\end{figure}

Figure~\ref{Fig_fiber} shows a schematic of a TOF with three sections exhibiting different propagation properties: the nanofiber waist (a) surrounded by the two taper transitions (b) and the original unstretched fiber (c). The intensitiy distribution of the corresponding guided mode is schematically represented by filled curves.

In the unmodified part of the single mode TOF, the light is guided inside the core via total internal reflection at the core-cladding boundary. Due to the low refractive index difference between core and cladding ($n_{\text{core}}/n_{\text{cladding}} \approx 1.0035$), this section is often designated as weakly guiding and Maxwell's equations are approximately solved by linearly polarized (LP) modes \cite{yariv}. In the taper transition, both cladding and core shrink. The decrease of the core diameter leads to a slight compression of the guided mode until the mode diameter reaches a minimum. Further reduction of the core diameter leads to an expansion of the mode into the cladding. The cladding then acts as the new guiding medium. In order to minimize transmission losses, this mode conversion should be adiabatic, i.e., there should be no coupling of the fundamental mode to higher transverse or radiative modes. This is achieved by maintaining shallow taper angles of a few milliradians \cite{love,love2}. 

Further decrease of the fiber diameter to the waist value causes the mode to be more and more laterally confined. The high refractive index difference at the cladding-air interface ($n_{\text{cladding}} / n_{\text{air}} \approx 1.5$) allows a much tighter radial confinement. Throughout the nanofiber waist, the diameter of the fiber remains constant in the sub-wavelength regime, and the guided mode exhibits tight radial confinement and strong evanescent field. At the second taper transition, the mode transformation is reversed. Note, that through this process the weakly guided LP$_{01}$ mode of the unstretched fiber is adiabatically transformed into the strongly guided HE$_{11}$ mode of the nanofiber and back. 

\begin{figure}[h]
\centering
  \resizebox{0.48\textwidth}{!}{%
  \includegraphics{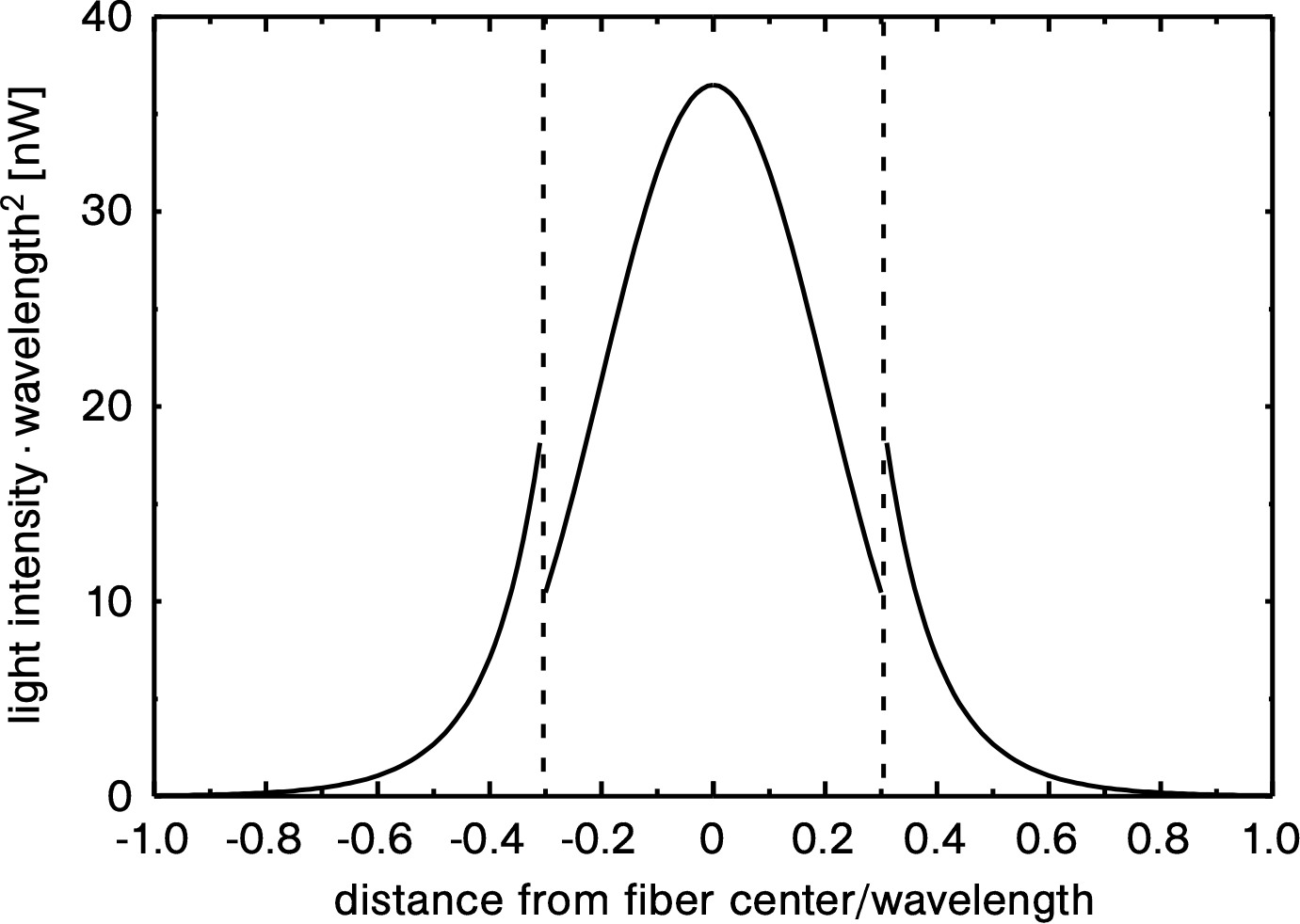}
}
    \caption{Radial intensity distribution in units of $1/\lambda^2$ of the fundamental mode HE$_{11}$ guided through a nanofiber as a function of the distance from the fiber center in units of the wavelength. Calculated according to \cite{le kien} for circularly or unpolarized light, a guided power of 10~nW, $d/\lambda=0.6$ and a refractive index $n=1.46$.}\label{Fig_int_distribution}
\end{figure}

The intensity distribution of a circularly or unpolarized HE$_{11}$ mode is plotted in Fig.~\ref{Fig_int_distribution} as a function of the distance from the fiber center. The intensity is maximal at the fiber center and decreases towards the surface. Due to the high refractive index jump at the surface, the radial component of the electric field exhibits a correspondingly strong discontinuity, while the tangential components of the electric field remain continuous \cite{le kien}. In the nanofiber, for visible wavelengths the decay length of the evanescent field is typically on the order of few hundreds of nanometers. The strong intensity achieved at the fiber surface provides excellent conditions for high-sensitivity spectroscopic measurements and non-linear optical experiments.

\newcommand{\figaeff}{Fig.~\ref{sensitivity}} % reference to Ruth's figure with A_eff

\begin{figure}[h]
\centering
  \resizebox{0.48\textwidth}{!}{%
  \includegraphics{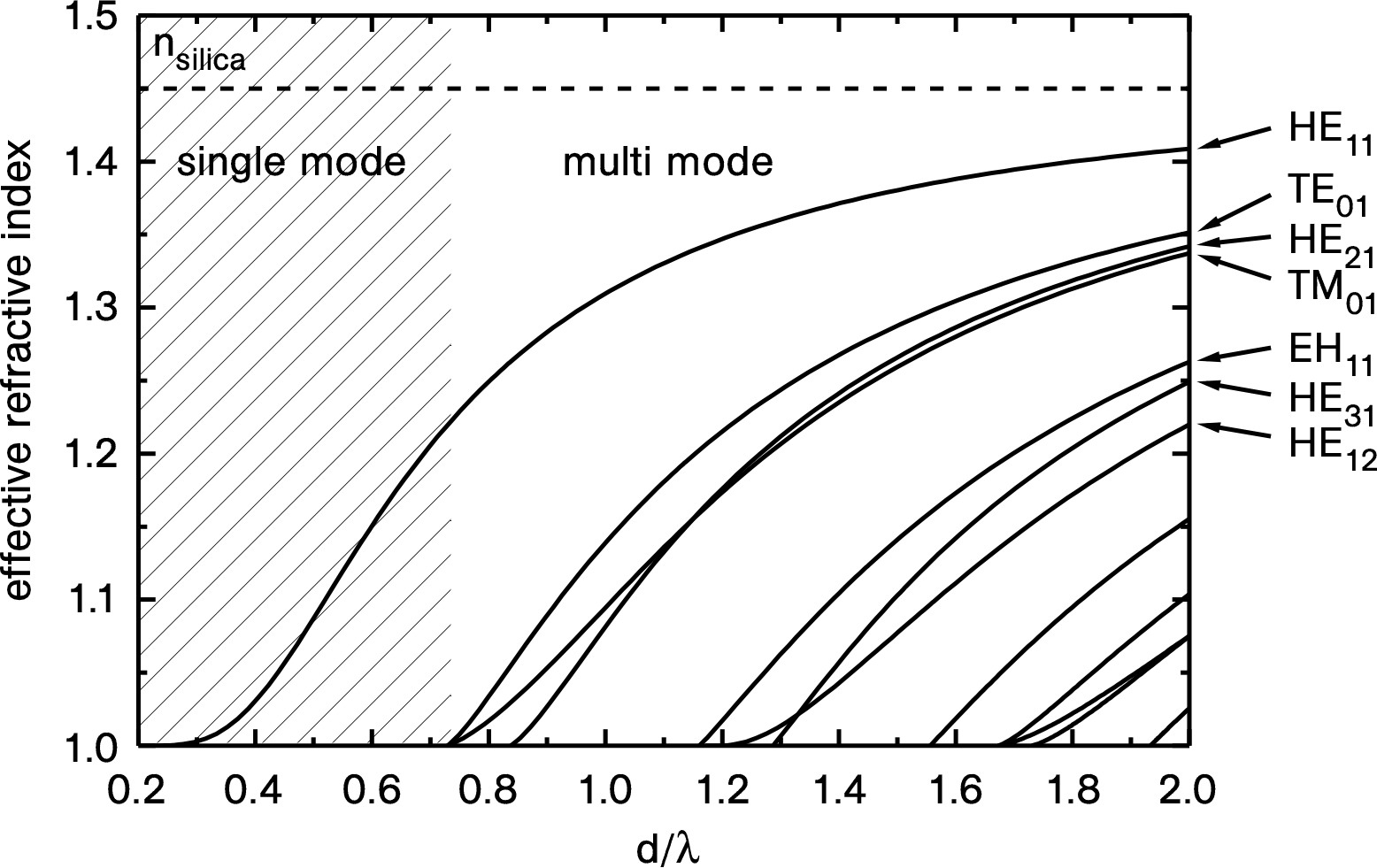}
}
    \caption{Effective refractive index for the fundamental mode (leftmost) and a few higher order modes
as a function of the fiber diameter in units of wavelength. The calculation was performed assuming a refractive index of $n_{\text{silica}} = 1.45$. Reproduced from \cite{shg}.}\label{fig:bonn_modes}
\end{figure}

Light propagation in a nanofiber is usually described using the concept of an effective refractive index $n_\text{eff} = c/v_\text{phase} = \beta/k_0$, where $c$ is the vacuum speed of light, $v_\text{phase}$ is the phase velocity of the mode, $\beta$ is the mode propagation constant, and $k_0 = 2\pi/\lambda$ is the vacuum wavenumber \cite{snyder,yariv,buck,meschede}. The propagation constant $\beta$ is obtained by solving the transcendental equation arising from the Maxwell's equations in conjunction with the boundary conditions applied to the nanofiber geometry \cite{yariv}. For this purpose, we have created a Matlab toolbox that allows us to automatically calculate all relevant parameters of the nanofiber guided modes \cite{oft}. As an example, Fig.~\ref{fig:bonn_modes} shows the effective refractive indices of the lowest nanofiber modes as a function of the waist diameter. The single-mode and multi-mode regime can be clearly distinguished in the figure. The cutoff value at which only the fundamental HE$_{11}$ mode propagates is wavelength-dependent and corresponds to $d/\lambda \approx 0.73$ for a refractive index of $n_{\text{silica}}=1.45$. As the fiber diameter decreases, a larger and larger fraction of the mode propagates outside of the glass, and the effective refractive index of the mode decreases to unity. On the other hand, a fiber diameter increase leads to a mode guided mostly entirely in the bulk and the effective refractive index approaches $n_{\text{silica}}$. Throughout this paper, with the exception of Sect.~\ref{sec:interferometer}, adiabatic tapers are used, i.e., to a good approximation the light is guided exclusively in the fundamental HE$_{11}$ mode.

% ==================== 3. Characterization ========================
\section{Characterization of the diameter profile}
\label{sec:characterization}

The knowledge of the fiber diameter profile is essential for a quantitative understanding of the light propagation and of nonlinear effects in nanofibers. In order to characterize the predetermined diameter profile used by the fiber pulling rig we performed scanning electron microscope (SEM) measurements. Although this method can offer an accuracy of 3$\%$, it is time-consuming and doesn't allow one to re-use the TOF because it has to be put onto a substrate and needs to be gold-coated \cite{shg,brambilla3}. Another method for measuring the nanofiber uniformity with a precission of 2-3$\%$ has been proposed \cite{birks2,sumetsky2} that, however, gives no information about the absolute fiber diameter. We have developed a non-destructive optical method for diameter measurements with an acurracy better than 2$\%$.

\subsection{Scanning electron microscope measurement}

\begin{figure}[h]
\centering
  \resizebox{0.48\textwidth}{!}{%
  \includegraphics{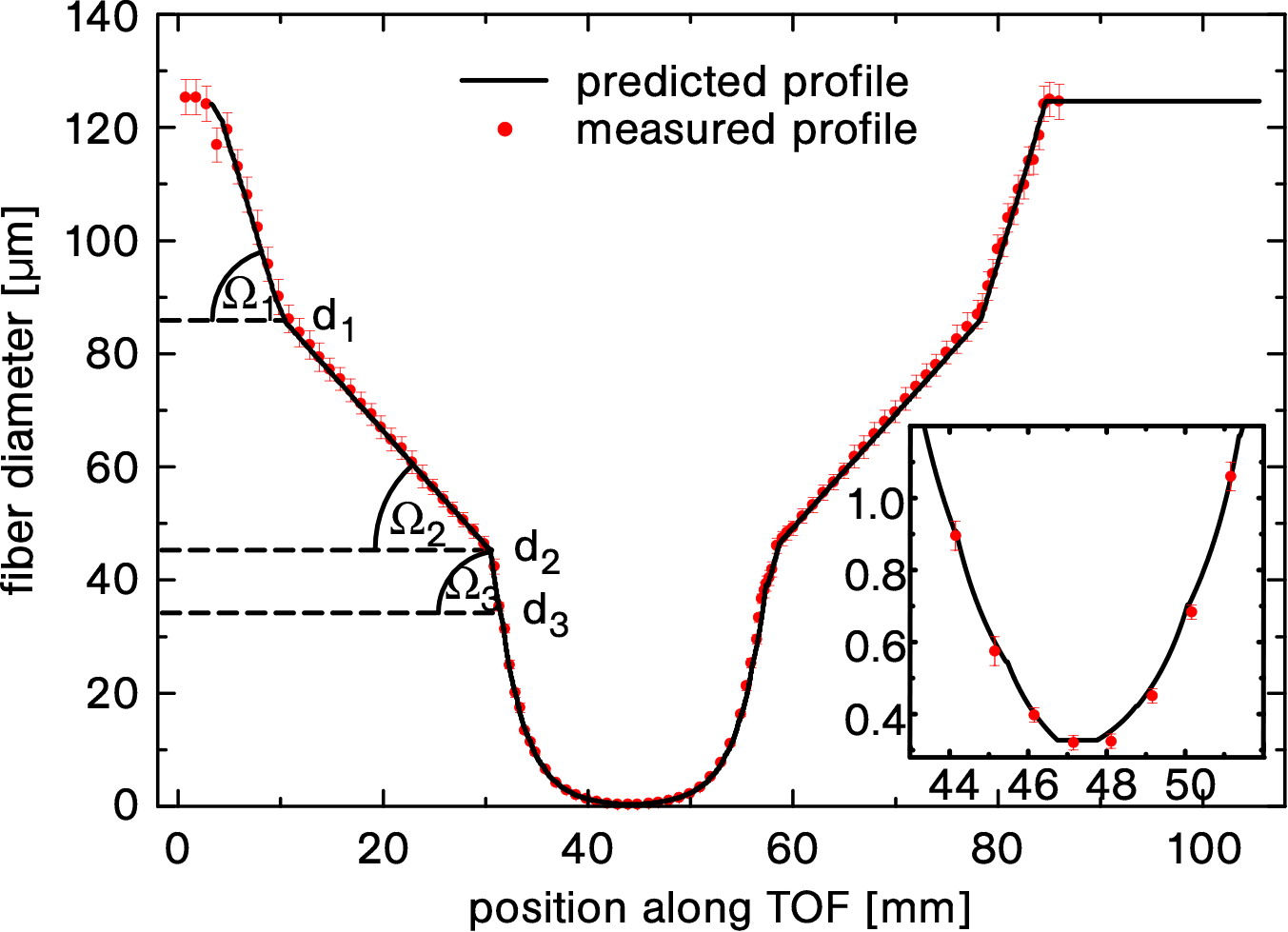}
}
    \caption{Comparison between a SEM measurement (points) and the predicted diameter profile (solid line) of a TOF with a nanofiber waist. Reproduced from \cite{transmission}.}\label{Fig1_transmission}
\end{figure}

Figure~\ref{Fig1_transmission} shows the diameter profile of a TOF with a 320-nm diameter nanofiber waist of 1-mm length measured with a SEM. It exhibits a three-fold linear taper transition with local taper angles of $\Omega_1 = 3$~mrad, $\Omega_2 = 1$~mrad and $\Omega_3 = 5$~mrad with the slope changing at the fiber diameters $d_1 \approx 84~\mu$m and $d_2 \approx 40~\mu$m. The solid line indicates the fiber diameter profile as predicted by our algorithm and the inset shows a magnification of the submicron-diameter region of the profile containing the waist with a length of 1~mm. The nanofiber was mounted on a gold coated substrate for the SEM measurement (see \cite{transmission} for technical details). In order to determine the local fiber diameter, we successively took electron micrographs along the TOF with a step size varying between 0.25 and 1~mm. The measured diameter profile closely follows the predicted profile with deviations smaller than 10$\%$. For the two linear sections with local taper angles $\Omega_1$ and $\Omega_2$, they even fall below 5$\%$. From this excellent agreement we conclude that our production procedure for nanofibers
indeed realizes the predicted profiles to a very good approximation.

\begin{figure}[h]
\centering
  \resizebox{0.48\textwidth}{!}{%
  \includegraphics{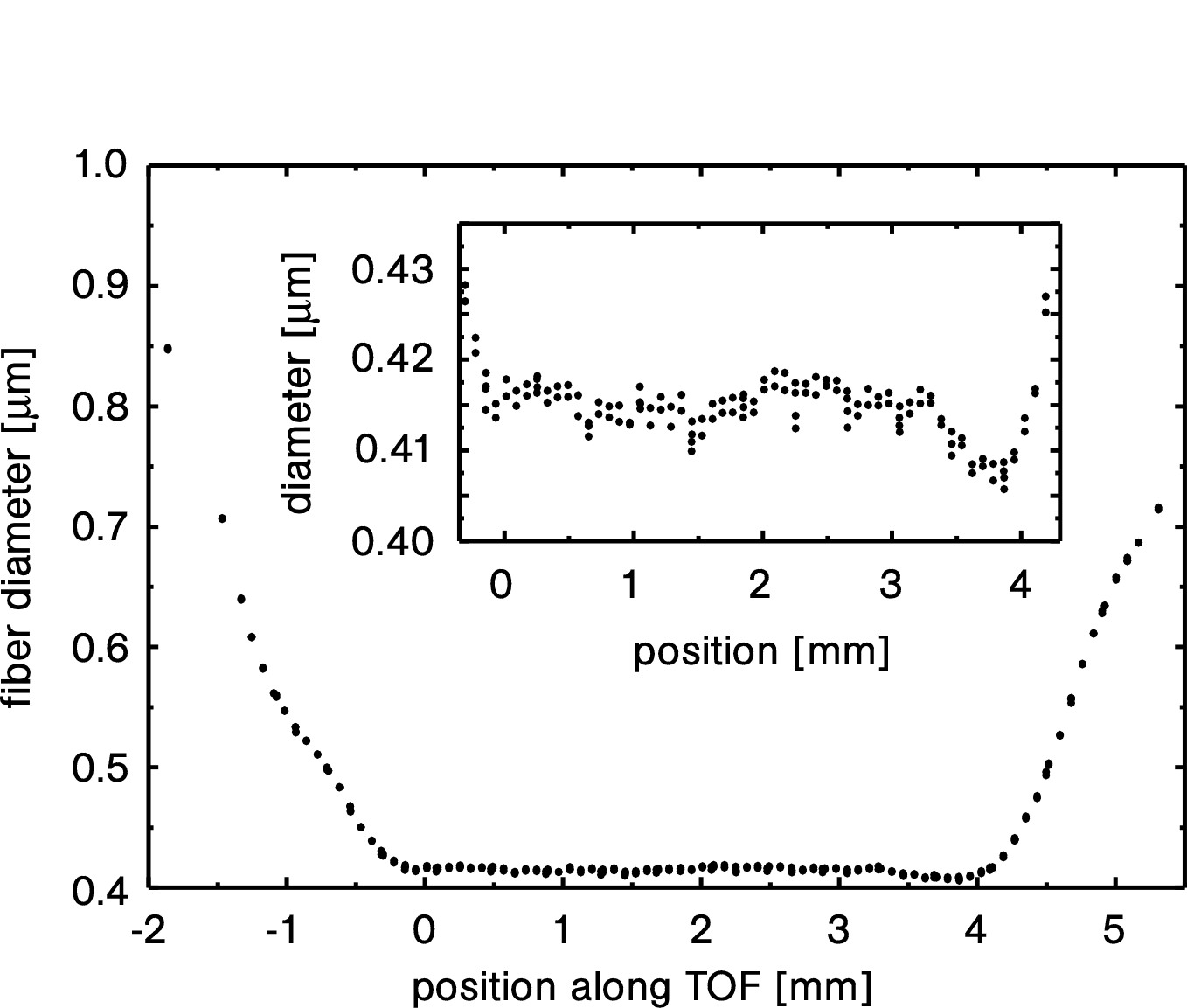}
}
    \caption{SEM measurement of the nanofiber-waist region of a TOF. The inset shows a vertically enlarged picture of the waist. Each point has been measured with an accuracy of better than 7~nm. Reproduced from \cite{shg}.}\label{fig:waist_sem}
\end{figure}

% ====================== Bonn_fibre_measurement =========================
Figure~\ref{fig:waist_sem} shows the detailed measurement of the waist region of a nanofiber using a Zeiss SUPRA~55 field-emission SEM. Since the manufacturer guaranties an accuracy of only $\pm5\percent$, we have recalibrated the SEM for each measurement series and thereby achieved an improvement of the absolute accuracy to below $\pm2\percent$ (see \cite{shg} for further technical details). Our fiber pulling rig produces a waist with less than $\pm1.5\percent$ diameter variation, sufficiently uniform for the experiments reported here. At the same time, the pulling algorithm predicted a waist diameter of 393~nm while the SEM measurement yield a value of 415~nm. This accuracy of about 5\percent agrees well with the global nanofiber shape measurement shown in Fig.~\ref{Fig1_transmission}.

% ======================== Optical measurement ======================
\subsection{Optical measurement}
\label{sec:bonn_diameter_measurement}

The optical method for measuring the waist diameter is based on second and third harmonic generation (SHG and THG) in a bare silica nanofiber in air \cite{grubsky2007,laegsgaard2010}. Efficient harmonic generation requires that the phase-matching condition is fulfilled, i.e., the effective refractive indices of the fundamental and harmonic waves have to be equal. Phase-matching for the two fundamental modes of both waves is difficult to achieve due to material dispersion and the wavelength dependence of the effective indices of the modes in a nanofiber (compare Fig.~\ref{fig:bonn_modes}). However, the same effective refractive index dependence on the fiber diameter allows one to fulfill the phase-matching condition if the harmonic light propagates in a higher transverse mode. If the wavelength changes, the effective refractive index curves shift and the phase-matching condition occurs at a different fiber diameter. There exists a one to one relation between the fiber diameter and the phase-matched wavelength for each mode.  

\begin{figure}[h]
\centering
  \resizebox{0.48\textwidth}{!}{%
  \includegraphics{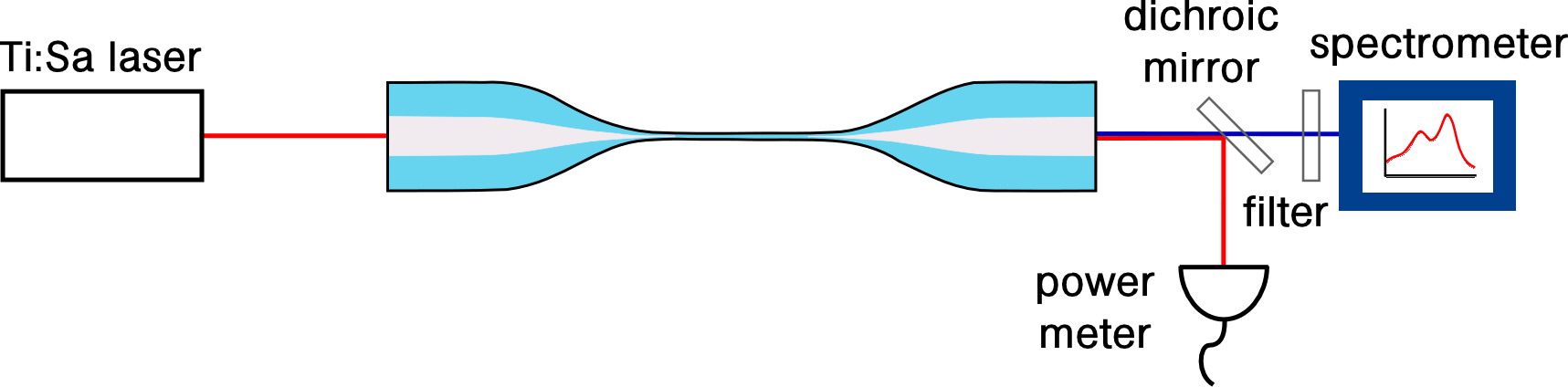}
}
    \caption{Scheme of the experimental setup for SHG and THG measurement.}\label{fig:fibre_measurement_optical_setup}
\end{figure}

Figure~\ref{fig:fibre_measurement_optical_setup} shows a schematic of the experimental setup. Light from a tunable pulsed Ti:Sa laser (1~ps pulse duration, 80~MHz repetition rate, 100~mW average power) is coupled into one end of the TOF and its wavelength is scanned in order to detect the SHG and THG peaks. A dichroic mirror at the output of the TOF reflects the infrared fundamental light, which is measured with a power meter. The transmitted SHG and THG light is analyzed by a spectrometer (see \cite{shg} for further details).

\begin{figure}[h]
\centering
  \resizebox{0.48\textwidth}{!}{%
  \includegraphics{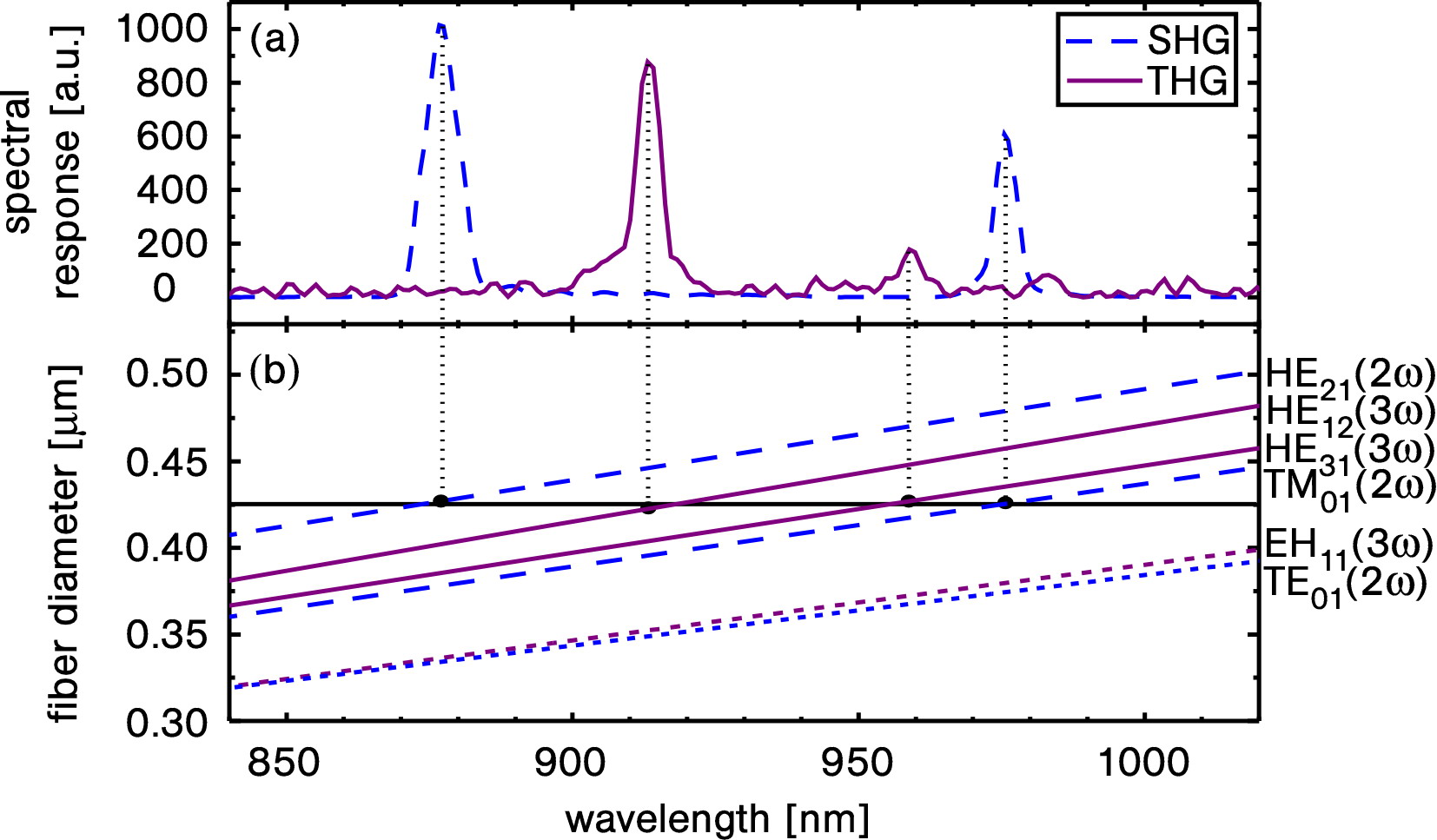}
}
    \caption{(a) The spectral response of a nanofiber for SHG (dashed line) and THG (solid line) plotted vs.~the fundamental wavelength. The spectrometer response of the THG measurement is scaled up for visibility. The four peaks correspond to phase matching of HE$_{11} (\omega)$ to the following modes (from left to
right): HE$_{21}(2\omega)$, HE$_{12}(3\omega)$, HE$_{31}(3\omega)$, TM$_{01}(2\omega)$. (b) Wavelength-dependent phase-matching diameter. Dashed lines: SHG, solid lines: THG, dotted lines: modes not observed due to lack of effective mode overlap with the fundamental mode. The horizontal line indicates the diameter of the investigated sample determined by this method. Reproduced from \cite{shg}.}\label{fig:diameter_vs_spectrum}
\end{figure}

Figure~\ref{fig:diameter_vs_spectrum} (a) shows the measured full response of the SHG and THG of a nanofiber. In Fig.~\ref{fig:diameter_vs_spectrum} (b), the phase-matching condition for harmonic generation is plotted. The nanofiber diameter is determined from the position of the SHG and THG peaks. 

For the sample measured in Fig.~\ref{fig:diameter_vs_spectrum}, the fiber waist diameter is obtained as 425.7~nm with a resolution of $0.5 \times 10^{-3}$ and an accuracy of $<2\%$. Taking into account the transparency window of silica, our method allows to measure fiber waists with diameters down to 190~nm and up to 1~$\mu$m.

% ====================== 4. Applications ==========================
\section{Applications in surface spectroscopy}
\label{sec:surface}

This section demonstrates surface spectroscopy of different types of nanofiber-adsorbed emitters. Further, we quantitatively analyze the experimental results.

\subsection{Experimental setup}
\label{sec:setup}

\begin{figure}[h]
\centering
  \resizebox{0.48\textwidth}{!}{%
  \includegraphics{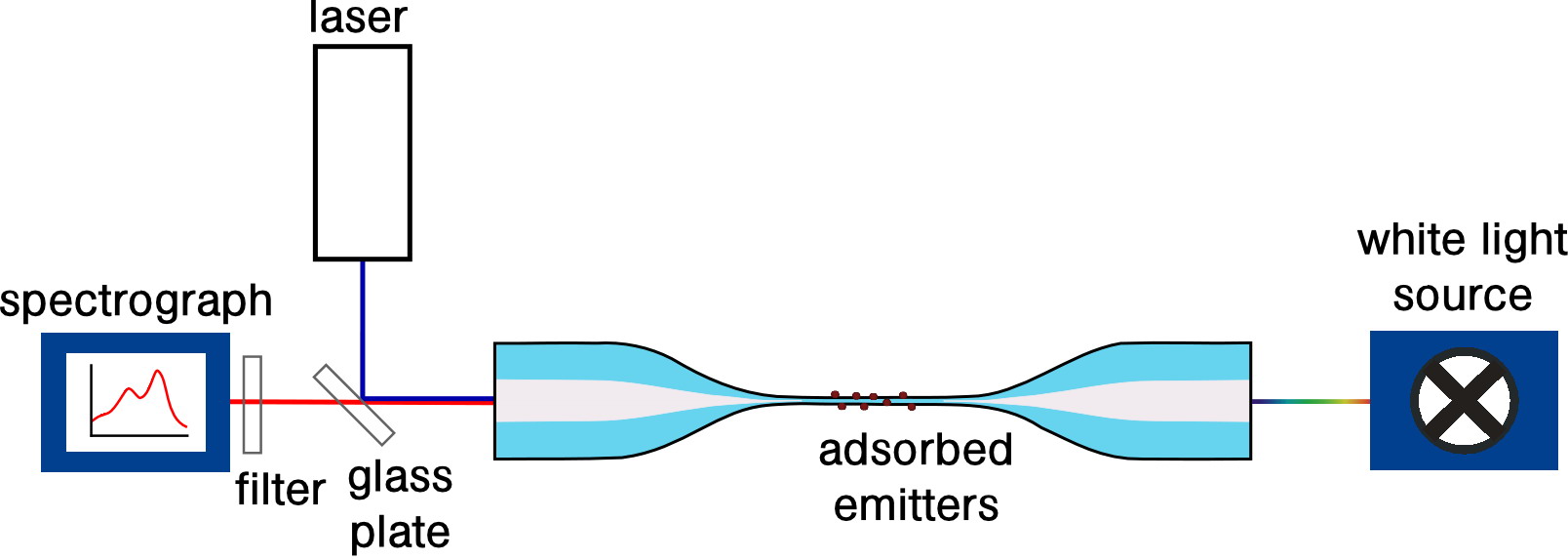}
}
    \caption{Scheme of the experimental setup for absorption and fluorescence spectroscopy.}\label{Fig1_fluorescence}
\end{figure}

Figure~\ref{Fig1_fluorescence} shows the schematic experimental setup for fiber-based absorption and fluorescence
spectroscopy. Depending on the species under study, the particles are deposited onto the nanofiber waist using different approaches, e.g., vapor deposition or immersion in a solution. The absorption of the deposited particles is measured using a conventional absorption spectroscopy configuration with a tungsten white light source and a spectrograph. Fluorescence spectra are obtained upon excitation by laser light coupled into the TOF after reflection from a glass plate. The counterpropagating fluorescence is collected after transmission through the glass plate by the same spectrograph as for the transmitted white light. This configuration avoids direct incidence of the excitation laser light into the spectrograph. The residual backscattered and reflected excitation laser light is suppressed by means of a razor edge filter, placed directly in front of the spectrograph's entrance slit.

% \clearpage

\subsection{Theoretical considerations} 
\label{theory}

In the following, we derive expressions for the measured absorption and fluorescence signals to be expected from the set-up described above. 

We consider a nanofiber with diameter $d$ and length $L$ covered with $n$ emitters, yielding a
surface coverage, i.e., number of emitters per surface area, of $\theta = n/\pi dL$. The absorbance
of light at a wavelength $\lambda$ propagating in the fundamental mode of this nanofiber is given
by $\eta(\lambda) = -\text{lg} \left( P_{\text{sig}}(\lambda)/P_{\text{ref}}(\lambda) \right)$, where $P_{\text{sig}}(\lambda)$ and $P_{\text{ref}}(\lambda)$ are the transmitted powers at the excitation wavelength $\lambda$ in the presence and absence of emitters, respectively. According to \cite{absorption}, the absorbance $\eta(\lambda)$ can be approximated as

\begin{equation}
	\eta (\lambda) \approx \frac{n \sigma (\lambda)}{\text{ln}(10)A_{\text{eff}} (\lambda)} = \frac{\theta \sigma (\lambda)}{\text{ln}10} \cdot \frac{ \pi d L}{A_{\text{eff}} (\lambda)},
\label{abs}
\end{equation}
if the absorption cross section of the emitter $\sigma (\lambda)$ is much smaller than the effective area of the guided fiber mode $A_{\text{eff}} (\lambda)$ which is defined by

\begin{equation}
	A_{\text{eff}} (\lambda) = \frac{P_{\text{ref}}(\lambda)}{I_{\text{surf}}(\lambda)}.
\label{aeff}
\end{equation}
$I_{\text{surf}}(\lambda)$ denotes the intensity of the evanescent field components at the fiber surface that actually couple to the emitters. $I_{\text{surf}}(\lambda)$ does account for a possible preferential orientation of the transition dipole of the emitter which might occur, e.g., when planar molecules lie flat at the fiber surface \cite{ogawa}. In this case, their transition dipole moment is oriented perpendicular to the radial electric field component of the guided fiber mode. Therefore, only the axial and azimuthal components of the electric field, calculated according to \cite{le kien}, contribute to $I_{\text{surf}}(\lambda)$. 

Equation~(\ref{abs}) can be rewritten as a function of the absorbance $\eta_{\text{free}}(\lambda) = \theta \sigma (\lambda) / \text{ln}10$ of a freely propagating beam \cite{absorption}

\begin{equation}
	\eta (\lambda) = \eta_{\text{free}}(\lambda) \xi (\lambda),
\end{equation}
where the factor $\xi (\lambda) =  \pi d L / A_{\text{eff}}$ gives the enhancement of the sensitivity of the fiber-based absorption measurement. With a length of the nanofiber $L$ on the order of millimeters and a diameter $d$ in the sub-micron range, we thus expect an increase by four orders of magnitude in the sensitivity in comparison with methods using freely propagating beams \cite{absorption}. 

\begin{figure}[h]
\centering
  \resizebox{0.48\textwidth}{!}{%
  \includegraphics{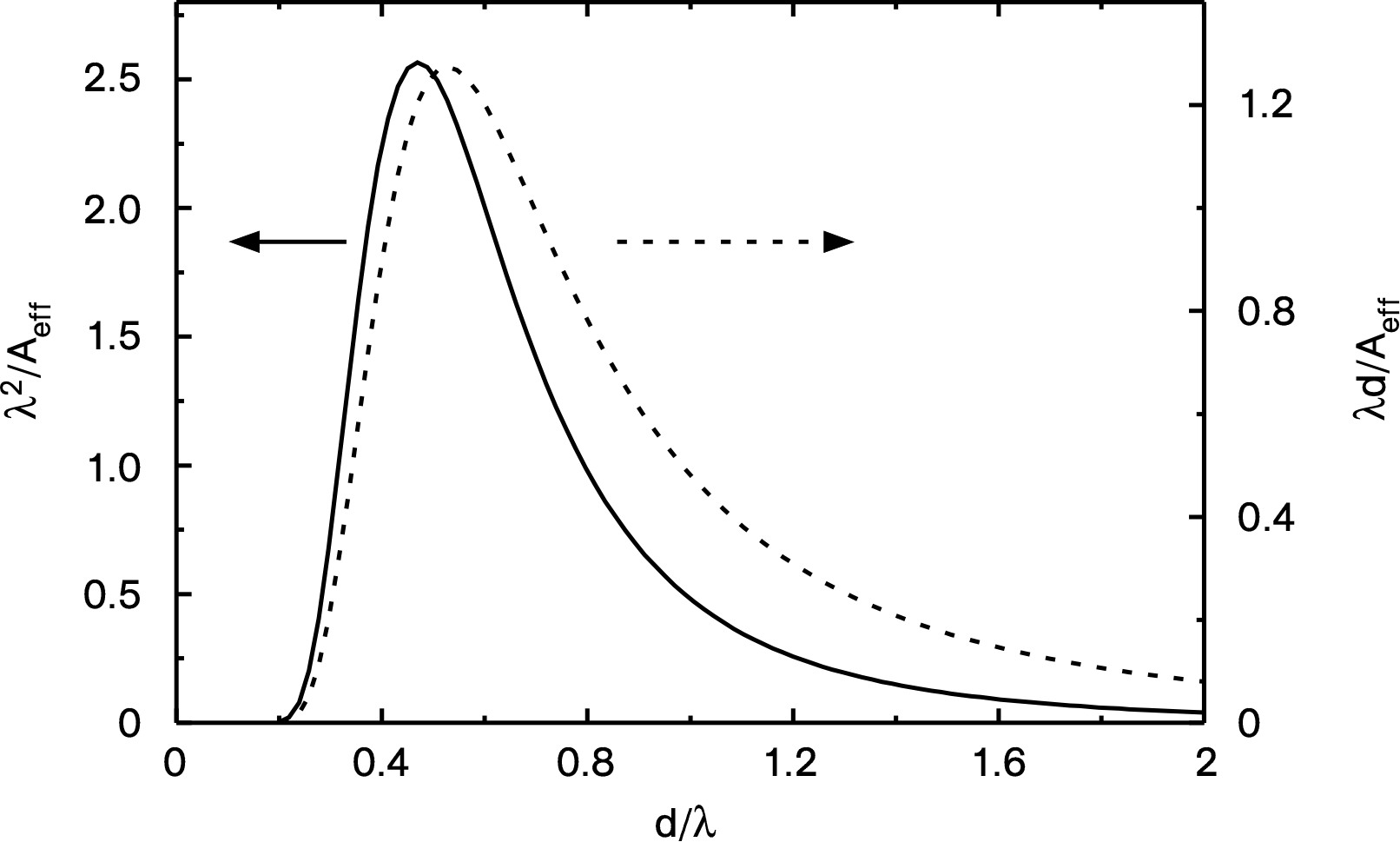}
}
    \caption{Plot of 1/$A_{\text{eff}}$ and $d/A_{\text{eff}}$ in units of $1/\lambda^2$ and $1/\lambda$, respectively, as a function of the fiber diameter $d$ in units of $\lambda$. $A_{\text{eff}}$ is calculated taking into account all three components of the electric field. Assuming a refractive index of 1.46 and ignoring material dispersion, both plots hold universally for any $\lambda$.}\label{sensitivity}
\end{figure}

The absorbance is directly proportional to the number of emitters $n$ and inversely proportional to the effective area $A_{\text{eff}}$. As can be seen from the second term in Eq.~(\ref{abs}), the quantity $ 1/ A_{\text{eff}}$ determines the sensitivity for a given number of emitters. It is plotted in Fig.~\ref{sensitivity} in units of $1/\lambda^2$ as a function of the fiber diameter and reaches a maximum for $d_{\text{max}} = 0.508 \cdot \lambda$. It rapidly goes to zero for small diameters because then the mode is only weakly bound and $A_{\text{eff}}$ diverges. The third term in Eq.~(\ref{abs}) shows that for spectroscopic measurements of thin layers, the quantity $d/A_{\text{eff}}$ should be maximized. This quantity is also plotted in Fig.~\ref{sensitivity} in units of $1/\lambda$ as a function of $d/\lambda$, reaching a maximum for  $d_{\text{max}} = 0.584 \cdot \lambda$.

In order to derive an expression for the fluorescence signal, we consider an emitter at position $z$ along the fiber waist that absorbs the fraction $\sigma (\lambda_{\text{exc}}) / A_{\text{eff}} (\lambda_{\text{exc}})$ of the power $P_{\text{sig}} (\lambda_{\text{exc}},z)$ of the excitation wavelength $\lambda_{\text{exc}}$. For the fiber diameter and the wavelengths considered in our experiments, about 20$\%$ of the fluorescence of a single dipole emitter on the fiber surface is expected to be coupled back into the fundamental guided mode of the fiber, 10$\%$ in each direction of the fiber \cite{le kien2}. As a result, the fluorescence emitted into the fiber mode by an emitter at a position $z$ along the fiber waist is proportional to the absorbed power:

\begin{equation}
	P_{\text{fl}} (\lambda,z) = C(\lambda) \cdot P_{\text{abs}} (\lambda_{\text{exc}},z),
	\label{fluo}
\end{equation}
where the proportionality factor $C(\lambda) \propto q(\lambda)/A_{\text{eff}} (\lambda)$ includes the wavelength dependent fluorescence quantum yield of the emitter $q(\lambda)$ and the average fractional emission
of the emitter into the guided fiber mode. The latter is an average over all possible orientations
of the emitter on the fiber surface. Furthermore, it is determined by the intensity of the
evanescent field at the fiber surface which is proportional to $1/A_{\text{eff}}$, see Eq.~(\ref{aeff}). Note, that the spectral overlap between the absorption and emission spectra in our system results in
a partial self-absorption of the emitted fluorescence by circumjacent emitters before it reaches
the output of the fiber. While the high sensitivity of our method allows us to perform measurements in a regime of low surface coverages, self-absorption is non-negligible at higher surface coverages. Detailed theoretical considerations in order to quantify this effect are presented in \cite{fluorescence}.

The expressions derived for the absorbance and fluorescence in Eq.~(\ref{abs}) and Eq.~(\ref{fluo}), respectively, are proportional to the wavelength dependent coupling to the fiber mode characterized by $1/A_{\text{eff}}$. The absorption and fluorescence spectra presented throughout this paper have been corrected accordingly.

\subsection{Organic molecules: linear excitation}

As a model system, we use 3,4,9,10-perylene-tetracarboxylic dianhydride molecules (PTCDA) because of their stability under evaporation at ambient conditions, their high quantum yield and the experimental and theoretical knowledge concerning their spectral
characteristics \cite{forrest}. Moreover, PTCDA molecules significantly change their spectral properties
depending on their arrangement on the surface \cite{proehl}. Therefore, these organic molecules are well suited for sensitivity studies.

The measurements were performed with a 320-nm diameter nanofiber with a length of 1~mm fabricated from a Nufern 460-HP fiber. This diameter ensures that only the fundamental mode is guided in the fiber waist for wavelengths longer than 450~nm, thereby matching the single mode cutoff wavelength of the unprocessed fiber.
Moreover, this diameter yields the maximum sensitivity for absorption spectroscopy in the
visible domain, as discussed in Sect.~\ref{theory}. The TOF used in this experiment has a transmission of up to 70$\%$ in the wavelength range between 450 and 650~nm. The molecules are deposited on the nanofiber waist by placing a crucible with PTCDA crystals below the fiber and by heating it to $325\,^{\circ}\mathrm{C}$. By convection, sublimated molecules are carried to the nanofiber waist where they are adsorbed. Interlaced measurements of absorption and fluorescence spectra are performed in order to determine the respective surface coverage (see Eq.~(\ref{abs})). The fluorescence spectra are obtained upon excitation with a freely running diode laser at a wavelength of 406~nm. The background signal induced by the excitation light in the fiber has been subtracted in the spectra. For further technical details, see \cite{fluorescence}.

\begin{figure}[h]
\centering
  \resizebox{0.48\textwidth}{!}{%
  \includegraphics{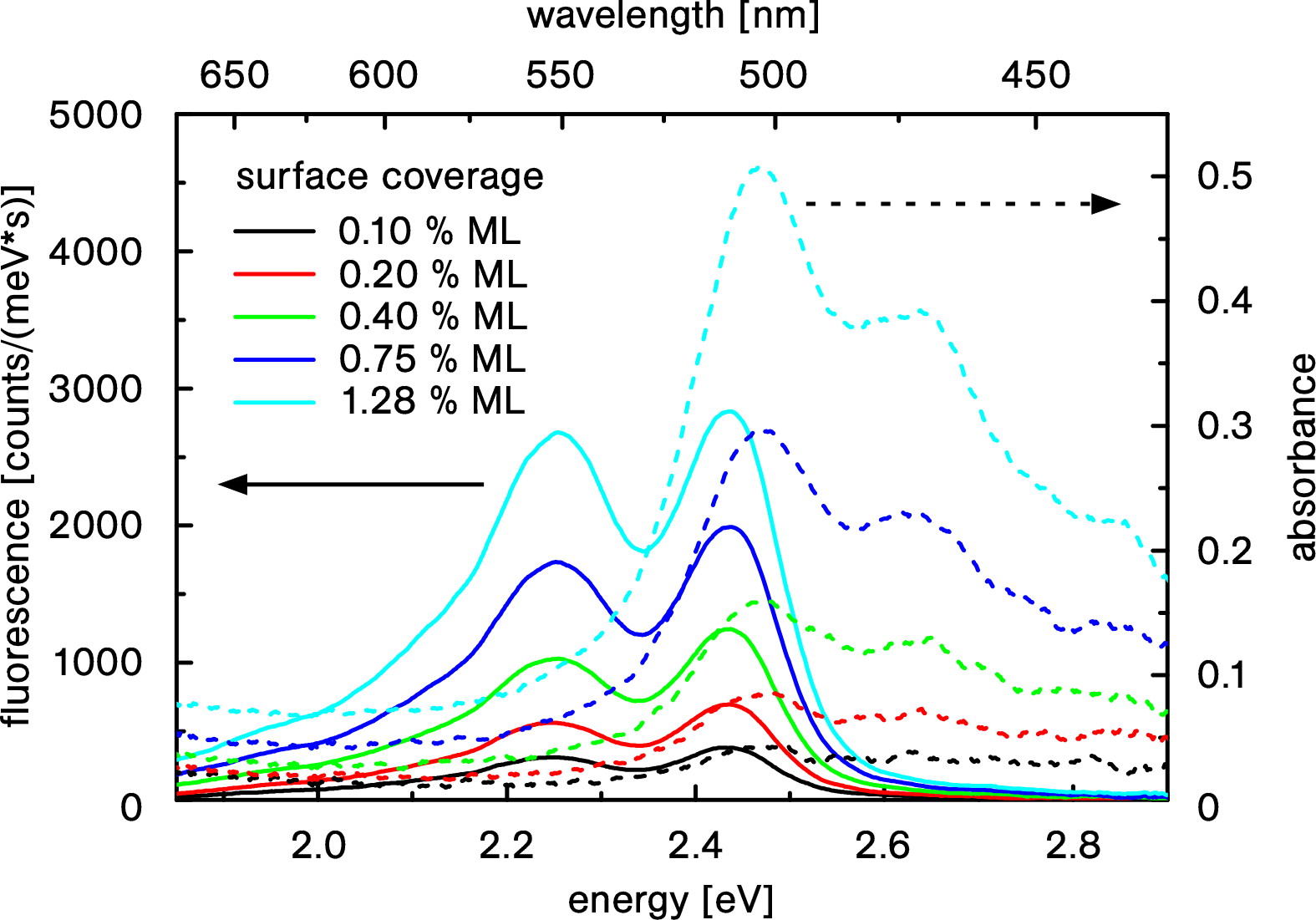}
}
    \caption{Fluorescence (left) and corresponding absorption (right) spectra of surface-adsorbed
PTCDA molecules during deposition. Five representative spectra within a range of surface
coverages between 0.10 and 1.28$\%$ of a ML are shown. Reproduced from \cite{fluorescence}.}\label{Fig2_fluorescence}
\end{figure}

Figure~\ref{Fig2_fluorescence} displays a series of \textit{in situ} fluorescence and absorption spectra recorded with about 8~$\mu$W laser power and about 3~nW white light power respectively. These powers are low enough not to saturate the molecules. For clarity, we chose five representative spectra taken during deposition of the molecules for different surface coverages. The shown absorption spectra are the mean of two consecutive spectra, one recorded 420 ms before and the other 420 ms after the corresponding fluorescence spectrum. Within this time interval, the deposition rate remains roughly constant, thus enabling us to determine the surface coverages underlying the measured fluorescence spectra by this averaging process. Using $\sigma= 4.1\times 10^{-16}$~cm$^2$ \cite{fluorescence} in Eq.~(\ref{abs}), we infer from the absorbance values ranging from 0.04 to 0.51 at the absorption maximum at 2.47~eV that 0.8$\times 10^6$ to 10.6$\times 10^6$ molecules cover the fiber waist. This corresponds to a surface coverage of 0.8$\times10^{11}$~cm$^{-2}$ to 10.6$\times10^{11}$~cm$^{-2}$ or 0.10$\%$ to 1.28$\%$ of a compact monolayer (ML) of flat lying PTCDA molecules, arranged in the herringbone structure of the (102) plane of the PTCDA bulk crystal \cite{forrest2}. The absorption spectra show a clear vibronic progression and agree well with differential reflectance spectroscopy (DRS) spectra of sub-monolayers of PTCDA on mica \cite{proehl}. The fluorescence spectra also show a vibronic progression as known for PTCDA in solution \cite{bulovic}. Note, that the absolute peak positions in both, absorption and fluorescence spectra, are slightly shifted in comparison with the literature due to the interaction of the adsorbed molecules with the fiber surface.

\begin{figure}[h]
\centering
  \resizebox{0.48\textwidth}{!}{%
  \includegraphics{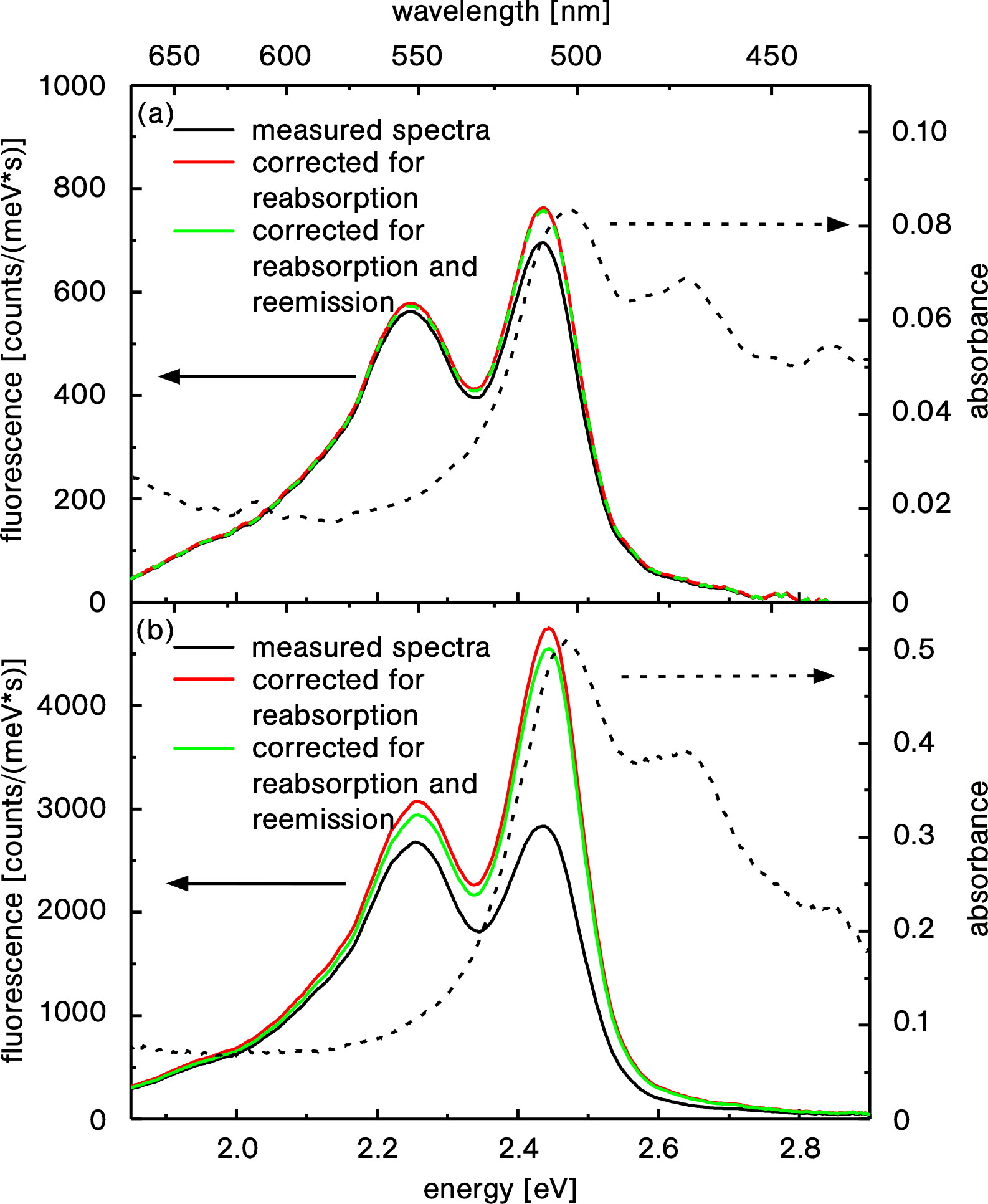}
}
    \caption{Influence of self-absorption and reemission on the measured fluorescence signal for
0.20$\%$ (a) and 0.75$\%$ (b) of a ML. For 0.75$\%$ of a ML, the signal corrected for self-absorption differs significantly from the measured spectrum. For 0.20$\%$ of a ML, the difference is only marginal. The correction for reemission does not have a significant influence on the spectral shape for neither of the two surface coverages. The absorbance for both surface coverages (dashed line) is shown for comparison of the spectral shape.}\label{Fig3_fluorescence}
\end{figure}

The mirror symmetry between absorption and fluorescence spectra that is characteristic for numerous molecular systems is retrieved after correction for the self-absorption effect. Figure~\ref{Fig3_fluorescence} displays two of the spectra from Fig.~\ref{Fig2_fluorescence} for about 0.75$\%$ and 0.20$\%$ of a ML and the ones corrected for self-absorption and further reemission of the reabsorbed fluorescence (see \cite{fluorescence} for further details). Note that, although for higher surface coverages like 0.75$\%$ of a ML the self-absorption changes the fluorescence spectra significantly, this effect is negligible for lower surfaces coverages like 0.20$\%$ of a ML. Summarizing, our sensitivity is high enough to perform studies for low surface coverages. At the same time, the fact that we are able to correct for self-absorption makes our method applicable for a large range of surface coverages.

Our method yields absorption and fluorescence spectra with a good signal to noise ratio for down to $\sim 10^6$ PTCDA molecules at room temperature. Under these conditions, thermal effects lead to strongly broadened molecular transitions with a FWHM of $\sim850$~cm$^{-1}$ or $\sim25$~THz \cite{scholz}. When cooling the fiber to cryogenic temperatures below 4~K, this thermal broadening could in principle be eliminated and $10^6$ times narrower lines of a few 10~MHz should be obtainable with organic dye molecules embedded in organic crystals \cite{ambrose}. Under these circumstances our method would thus also allow us to carry out spectroscopic studies on single molecules.

\begin{figure}[h]
\centering
  \resizebox{0.48\textwidth}{!}{%
  \includegraphics{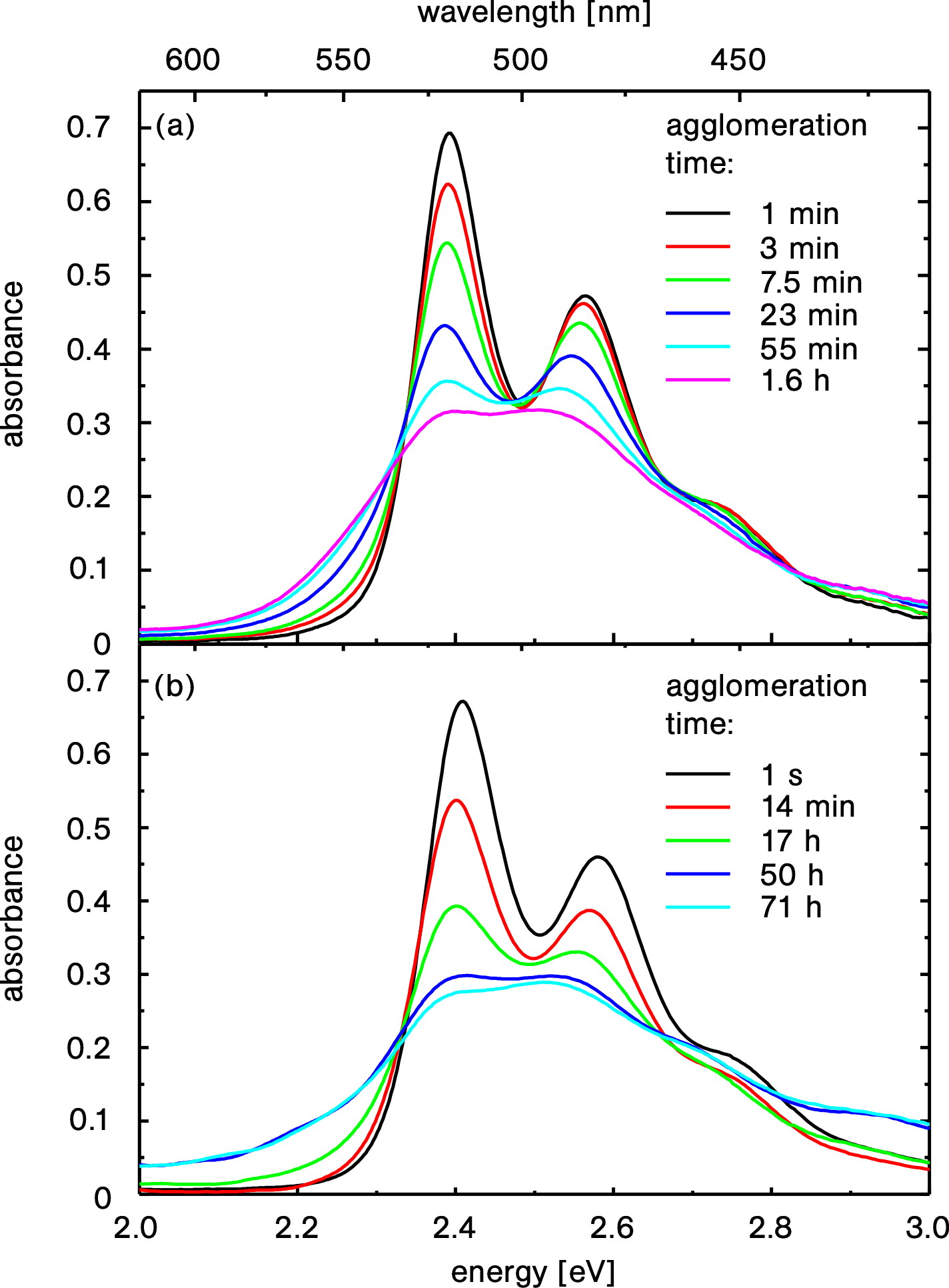}
}
    \caption{Time evolution of the spectral absorption of a constant number of molecules at ambient (a) and UHV (b) conditions, showing the agglomeration dynamics of the molecules on the fiber surface.}\label{Fig_olga}
\end{figure}

The high sensitivity of our method also allowed us to study the agglomeration dynamics of the adsorbed molecules on a second to minutes time scale at ambient conditions. Figure~\ref{Fig_olga} (a) shows absorption spectra measured during the ripening of the film after the molecular deposition has been stopped. We observe a continuous change of the shape of the spectra, evolving from monomer-like to oligomer-like. At the beginning of this process (up to 23~min), the absorbance remains constant at the energies of about 2.34~eV and 2.67~eV. These so-called isosbestic points result from the fact that, at these energies, the monomer and the dimer phase of PTCDA have the same molar absorptivity. This observation thus indicates that the total number of molecules remains nearly constant during the ripening. Additional kinetic details of the ripening process can be found in \cite{absorption}. Note that reordering of PTCDA molecules on a mica surface at ambient condition has already been observed. However, the measurement was not resolved in time \cite{proehl2}. The fast agglomeration process is attributed to the presence of water at the surface that increases the mobility of the surface-adsorbed molecules and thus accelerates their reordering. In order to desorb pollutants from the fiber surface, we set up an ultrahigh vacuum (UHV) chamber, allowing us to work with a more well defined surface. Figure~\ref{Fig_olga} (b) displays the absorption spectra measured during the agglomeration process in the UHV environment. A comparison with the analogue process at ambient conditions (Fig.~\ref{Fig_olga} (a)) shows qualitatively the same agglomeration dynamics. However, at UHV conditions, the process is slowed down by about two orders of magnitude, thereby showing that the mobility of the molecules in a vacuum environment is significantly reduced in comparison with ambient conditions. Both measurements have been performed with the same nanofiber diameter and the molecular deposition has been stopped at about the same surface coverage.

% ======================  Surface spectroscopy: TPF ==========================
% bonn_tpf
\subsection{Organic molecules: two-photon excitation}

The high light intensity and long interaction length (``the infinite focus'') makes nanofibers a suitable platform for nonlinear optical experiments. Here we present two-photon excited fluorescence (TPEF) measurements of organic molecules. The theory presented in Sect.~\ref{theory} holds for linear excitation and it has to be slightly modified for two-photon excitation. In particular, an emitter at position $z$ along the fiber waist absorbs now a fraction $\sigma_{2\text{Ph}} (\lambda_{\text{exc}}) / A_{\text{eff}} (\lambda_{\text{exc}})$ of the two-photon power $P_{\text{sig}}^2 (\lambda_{\text{exc}},z)$ of the excitation wavelength $\lambda_{\text{exc}}$, where  $\sigma_{2\text{Ph}} (\lambda_{\text{exc}})$ denotes the two-photon absorption cross section of the emitter.

We use the fluorescent dye Rhodamine 6G (Rh6G), which exhibits a high fluorescence quantum yield \cite{kubin1982}. Unlike PTCDA, most organic dye molecules undergo thermal decomposition by evaporation at atmospheric pressure. In order to extend the fiber surface spectroscopy to a larger variety of organic molecules, we developed a simple approach---the ``drip method''. The molecules are dissolved in a spectroscopic-grade solvent and a drop of this solution is dripped onto the nanofiber waist using a pipette. A thin film of the solution covers the fiber surface. Subsequently, the solvent evaporates and the molecules remain adsorbed at the fiber surface. 

With slight modifications, the experimental setup corresponds to the one shown in Fig.~\ref{Fig1_fluorescence}. For excitation, we use a pulsed Ti:Sa laser at 870~nm wavelength with a pulse duration of 1~ps, a repetition rate of 80~MHz and a $~10^4$-fold higher peak intensity compared to the corresponding continuous wave intensity. Further, in order to prevent laser power losses, the glass plate has been exchanged by a dichroic mirror that separates the infrared light of the Ti:Sa laser from the emitted visible light from the TPEF measurement. We use a 450-nm diameter nanofiber with a length of 5~mm fabricated from a Fibercore SM800 fiber.

\begin{figure}[h]
\centering
  \resizebox{0.48\textwidth}{!}{%
  \includegraphics{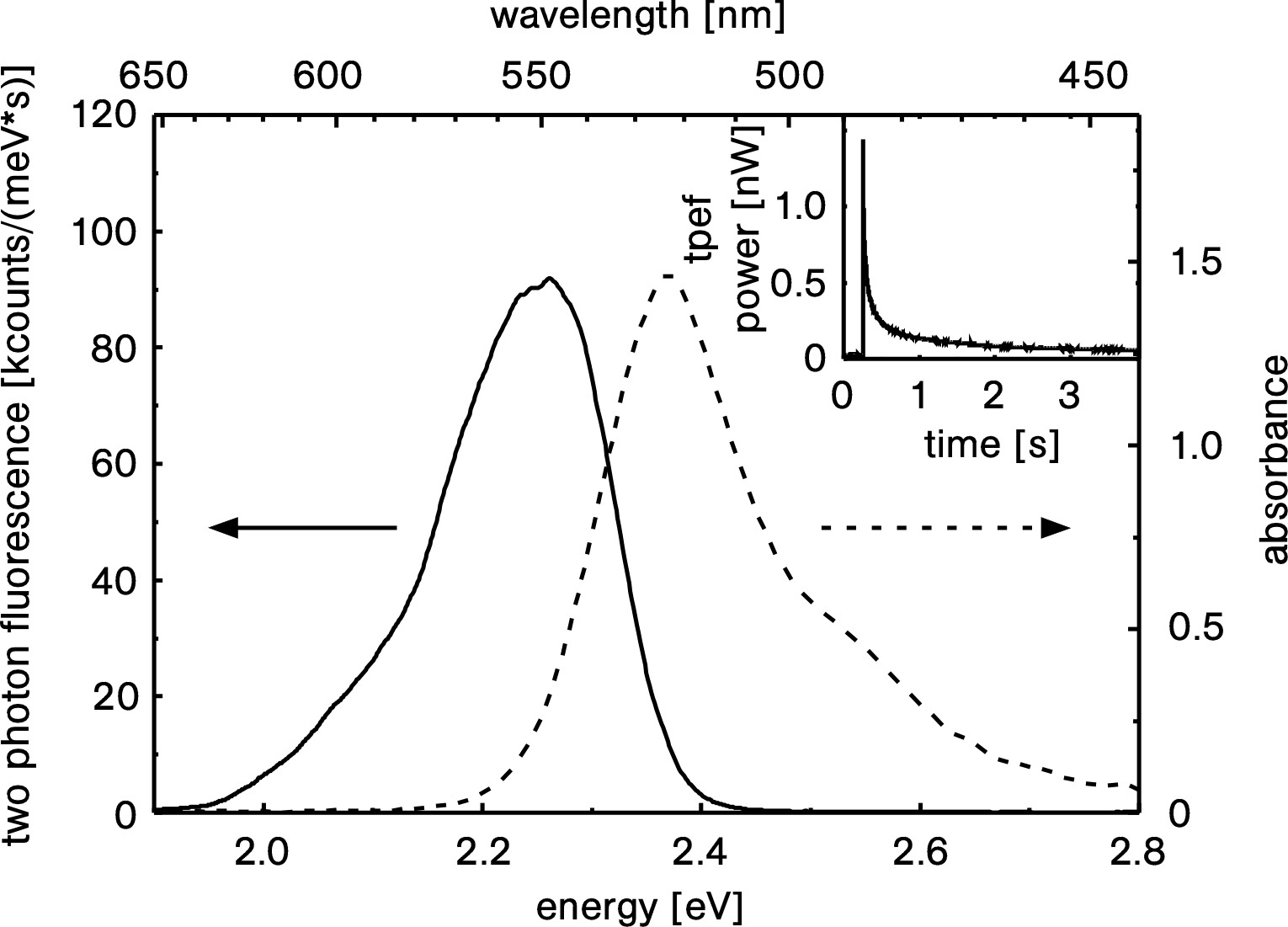}
}
    \caption{Two-photon excited fluorescence (solid line) and absorption (dashed line) spectrum of $4 \times 10^7$ Rhodamine 6G molecules (1.1~\% of a ML) adsorbed on a nanofiber. Inset: An independent measurement of the time evolution of the TPEF response, showing photobleaching of $2 \times 10^7$ adsorbed Rhodamine 6G molecules  (0.6~\% of a ML).}
    \label{fig:Fig_two_photon}
\end{figure}

The moleculess have been deposited with a 0.016~\% mass concentration solution of Rh6G in ethanol. The corresponding absorption spectrum of Rh6G adsorbed on the nanofiber is shown in Fig.~\ref{fig:Fig_two_photon}. Using the peak absorbance value of 1.47 and the absorption cross section of $6.7 \times 10^{-16}\text{~cm}^2$ at 2.37~eV (calculated according to \cite[ref.~23]{fluorescence} and \cite{du1998}), we estimate the number of Rh6G molecules adsorbed on the nanofiber waist to be $4 \times 10^7$. With the surface density of a ML of Rh6G molecules adsorbed on silica of $5\times 10^{13}\text{~cm}^{-2}$ \cite{heinz1982}, this corresponds to a surface coverage of 1.1~\% of a ML. By exciting the molecules with an average power of 20~mW of pulsed light, we detect the TPEF spectrum also plotted in Fig.~\ref{fig:Fig_two_photon}. It shows the typical mirror-image relation between the absorbance and fluorescence spectra \cite{birks1963}. The plotted fluorescence spectrum has been already corrected for the self-absorption effect. Both, absorption and fluorescence spectra, agree qualitatively well with the measurements performed in solution \cite{du1998}. However, there is a shift in the absolute peak positions which can be attributed to the interaction of the molecules with the fiber surface.

The TPEF spectrum is obtained directly after starting the excitation process since the TPEF intensity decreases rapidly due to photobleaching of the molecules. In fact, in a separate measurement, the temporal behavior of the TPEF strength has been recorded by replacing the spectrometer with a photomultiplier. The result is shown in the inset of Fig.~\ref{fig:Fig_two_photon}. Starting with an initial value of $2 \times 10^7$ adsorbed Rh6G molecules, the TPEF vanishes within seconds. At this point the absorbance is also zero and therefore, no undamaged molecules remain adsorbed at the fiber surface. 

% ====================== Surface spectroscopy: QD ==========================
% bonn_qd
\subsection{Quantum dot spectroscopy}

We have further investigated whether the spectroscopic method used for organic molecules can be also applied to quantum dots. Quantum dots are nanometer-scale semiconductor particles with a high fluorescence quantum yield, a much better resistance to photobleaching in comparison to organic molecules and a fluorescence emission wavelength that is size-dependent.

The experimental setup shown in Fig.~\ref{Fig1_fluorescence} has been used. For excitation, we use a continuous-wave diode laser at a wavelength of 405~nm. The measurement was performed using a 450-nm diameter nanofiber with a waist length of 5~mm fabricated from a Fibercore SM800 fiber.

\begin{figure}[h]
\centering
  \resizebox{0.48\textwidth}{!}{%
  \includegraphics{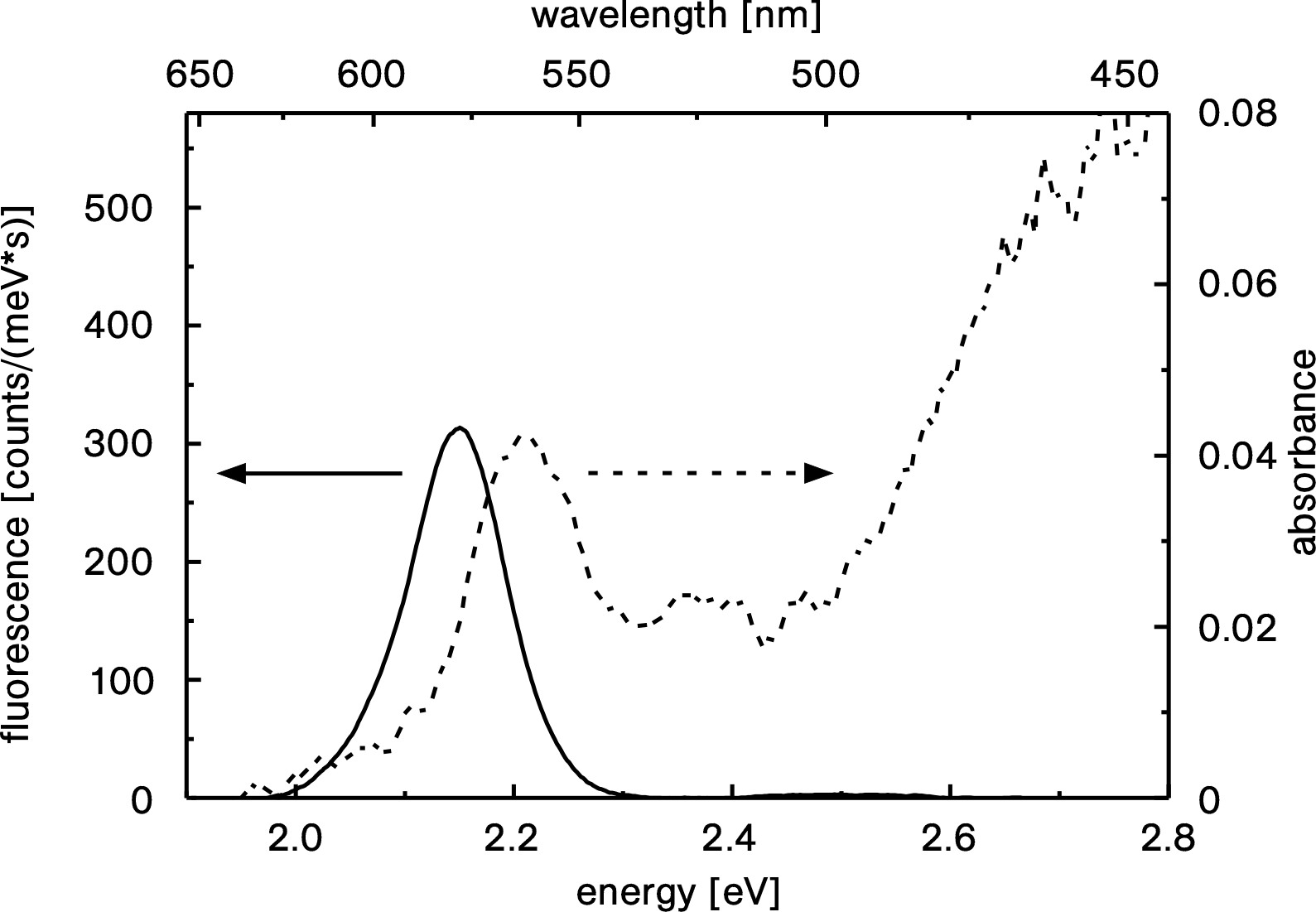}
}
    \caption{Fluorescence (solid line) and absorption (dashed line) spectrum of $3 \times 10^5$ CdSe quantum dots (0.3\% of a ML) adsorbed on an nanofiber. 
}\label{fig:Fig_qd}
\end{figure}

A solution of CdSe quantum dots (PL-QD-O-570 from Plasmachem GmbH, core diameter 3.3~nm, total diameter 8.9~nm) dissolved in heptane with a mass concentration of 0.036~\% was deposited by the drip method on the nanofiber waist. The absorption spectrum is shown in Fig.~\ref{fig:Fig_qd}. From the absorbance value of 0.043 and the absorption cross section of $8 \times 10^{-16}\text{~cm}^2$ at 2.22~eV (calculated from the molar extinction coefficient of $2.1\times 10^{5}\text{~M}^{-1}\text{~cm}^{-1}$ provided by the manufacturer), we estimate that about $3 \times 10^5$ quantum dots were adsorbed on the nanofiber waist. The corresponding surface coverage of 0.3~\% of a ML is calculated by assuming that the spherical quantum dots arrange in one full monolayer in the geometrically most compact form. The fluorescence spectrum, obtained by exciting the quantum dots with 36~nW of laser light at 405~nm, is also shown in Fig.~\ref{fig:Fig_qd}. The fluorescence maximum at 2.15~eV differs slightly from the manufacturer specifications of (2.175$\pm$0.020)~eV for quantum dots dissolved in chloroform. This discrepancy could be attributed to the interaction of the quantum dots with the nanofiber surface. Note that, for an absorbance as low as 0.043, the effect of self-absorption is negligible.

% ====================== Cs vapor ==========================
% bonn_cs_vapor
\section{Caesium vapor spectroscopy}
\label{sec:cs}

\newcommand{\degree}{\ensuremath{\,^\circ\text{C}}}
\newcommand{\D}[1]{\ensuremath{\text{D}_#1}}

The strong evanescent field surrounding the nanofiber waist of a TOF allows light-matter interaction not only with surface-adsorbed species but also with media surrounding the fiber. This feature allowed us to perform hot caesium vapor spectroscopy with a nanofiber.

The experimental setup consists of a tunable single-mode diode laser, a nanofiber in a vacuum chamber, and a silicon photodiode as a detector. The chamber contains a reservoir of liquid caesium, which is heated to 70\degree{}, thereby providing a caesium vapor pressure of approximately $10^{-4}$~mbar during the experiment. The rest of the chamber is heated to 80\degree{} in order to prevent caesium condensation anywhere outside the reservoir. The measurement was performed with a 450-nm diameter nanofiber with a length of 5~mm fabricated from a Fibercore SM800 fiber.

\begin{figure}[h]
\centering
  \resizebox{0.48\textwidth}{!}{%
  \includegraphics{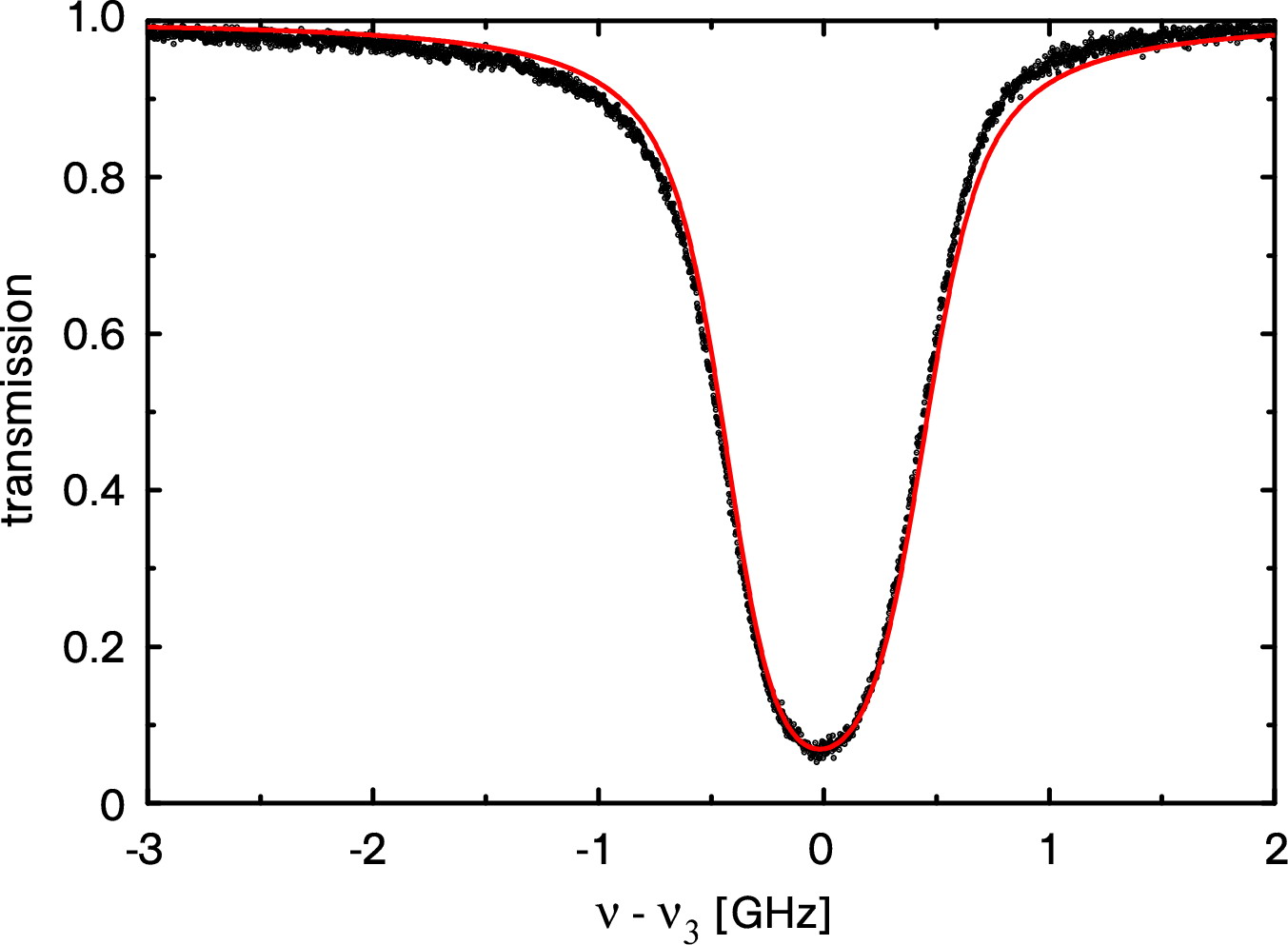}
}
    \caption{Measured (points) and fitted (solid line) caesium absorption spectra.}\label{fig:cs_transmission}
\end{figure}

Figure~\ref{fig:cs_transmission} shows the measured transmission through the nanofiber for an input laser power of 20~nW while scanning around the caesium \D2 resonance line at a wavelength of 852.3~nm. The data has been fitted using a function which is the sum of three Voigt profiles corresponding to the three hyperfine transitions ($F=3$ to $F'=2,\,3,\,4$) of the \D2 line: 

\begin{eqnarray}
	A(\nu) = \sum_{F'=2}^4 a_{F'} \int_{-\infty}^{\infty}  \frac{\exp\left(-x^2/(2\sigma^2)\right)}{\sigma\sqrt{2\pi}} \times \nonumber\\
	\times	\frac{2\gamma}{\pi((\nu- \nu_{F'}-x)^2+\gamma^2/4)}  
	 \text{d}x,
\label{cs}\end{eqnarray}
where $\nu$ is the frequency, $\nu_3 = 351.7$~THz the frequency of the $3\rightarrow3$ transition, $\nu_4 -\nu_3 = 201.3$~MHz and $\nu_3 - \nu_2 = 151.2$~MHz the separations of the hyperfine caesium levels \cite{steck2010}, $\gamma$ and $\sigma$ the widths of the Lorentzian and Gaussian components of the Voigt curve.  The hyperfine transition strength factors for the $3\rightarrow2$, $3\rightarrow3$, $3\rightarrow4$ transitions, have the values $a_2 = 5/14$, $a_3 = 3/8$, and $a_4 = 15/56$, respectively \cite{steck2010}.

The fit yields the values $\sigma=178$~MHz and $\gamma=130$~MHz. The Gaussian width agrees well with the theoretically expected value for Doppler broadening at a temperature of 80\degree:

\begin{equation}
	\sigma_\text{theoretical} = \nu_0\sqrt{\frac{k_\text{B}T}{mc^2}} = 174~\text{MHz},
\end{equation}
where $\nu_0=351.73$~THz is the frequency of the \D2 line, $k_\text{B}$ the Boltzman constant, $T = 353~\text{K}$ is the tempeature, and $m=2.21 \times 10^{-25}$~kg the mass of a caesium atom.

The inferred Lorentzian width is significantly larger than the natural line-width $\gamma_\text{lifetime}=1/(2\pi\tau)=5.2~\text{MHz}$, where $\tau$ is the \D2 lifetime of 30.4~ns. This may be explained by the fact that the thermal atoms pass the evanescent field of the fiber within a typical transit time of about 1~ns. 

These measurements show that nanofibers can also be used for high-resolution spectroscopic studies of gases and liquids surrounding the fiber waist. A similar experiment of nanofiber-based vapor spectroscopy has been done with rubidium vapor \cite{hendrickson}.

% ====================== Interferometer ==========================
% bonn_interferometer
\section{Towards dispersive spectroscopy}
\label{sec:interferometer}

Dispersive spectroscopy is interesting because it avoids destructive effects like e.g. molecule photobleaching. A Mach-Zehnder interferometer (MZI) is the natural candidate for this purpose. With the nanofibers we realize the MZI by modal interference.
The interference is achieved using the single-mode fiber pigtails of the TOF as the interferometer input and output, the down-taper as a beam splitter and the up-taper as a recombiner. In comparison with the nanofiber described above, the taper slope here is steeper, i.e. non-adiabatic, in order to induce coupling between the fundamental and higher order modes. If the entire nanofiber is cylindrically symmetric around its axis, light fields can be coupled only between the modes with the same azimuthal symmetry, e.g. from the HE$_{11}$ mode to HE$_{12}$, HE$_{13}$ etc. At the same time, a waist diameter can be chosen such that the HE$_{13}$ and the higher HE$_{1n}$ modes are cut off (see Fig.~\ref{fig:bonn_modes}), while the two lowest modes HE$_{11}$ and HE$_{12}$ are guided along the waist. Since the output pigtail is a single-mode fiber, the HE$_{12}$ mode is not guided there. The output power thus depends on the interference in the up-taper, determined by the relative phase accumulated by the HE$_{11}$ and HE$_{12}$ modes as they propagate along the waist: light can be either coupled back into the HE$_{11}$ mode and guided out, or to the HE$_{12}$ mode and be lost by coupling to radiative modes of the up-taper. Therefore the output power is given by:

\begin{equation}
 P_\text{out} \sim \sin (\varphi_{11} - \varphi_{12}), \quad\text{with}\quad \varphi_{1n} = \beta_{1n} L,
\end{equation} 
where $\beta = 2\pi n_\text{eff}/\lambda$ is the propagation constant of a mode, $L$ is the propagation distance, $\varphi_{1\text{n}}$ is the phase accumulated by the HE$_{1\text{n}}$ mode, and $n=1$ or~2.

\begin{figure}[h]
\centering
  \resizebox{0.48\textwidth}{!}{%
  \includegraphics{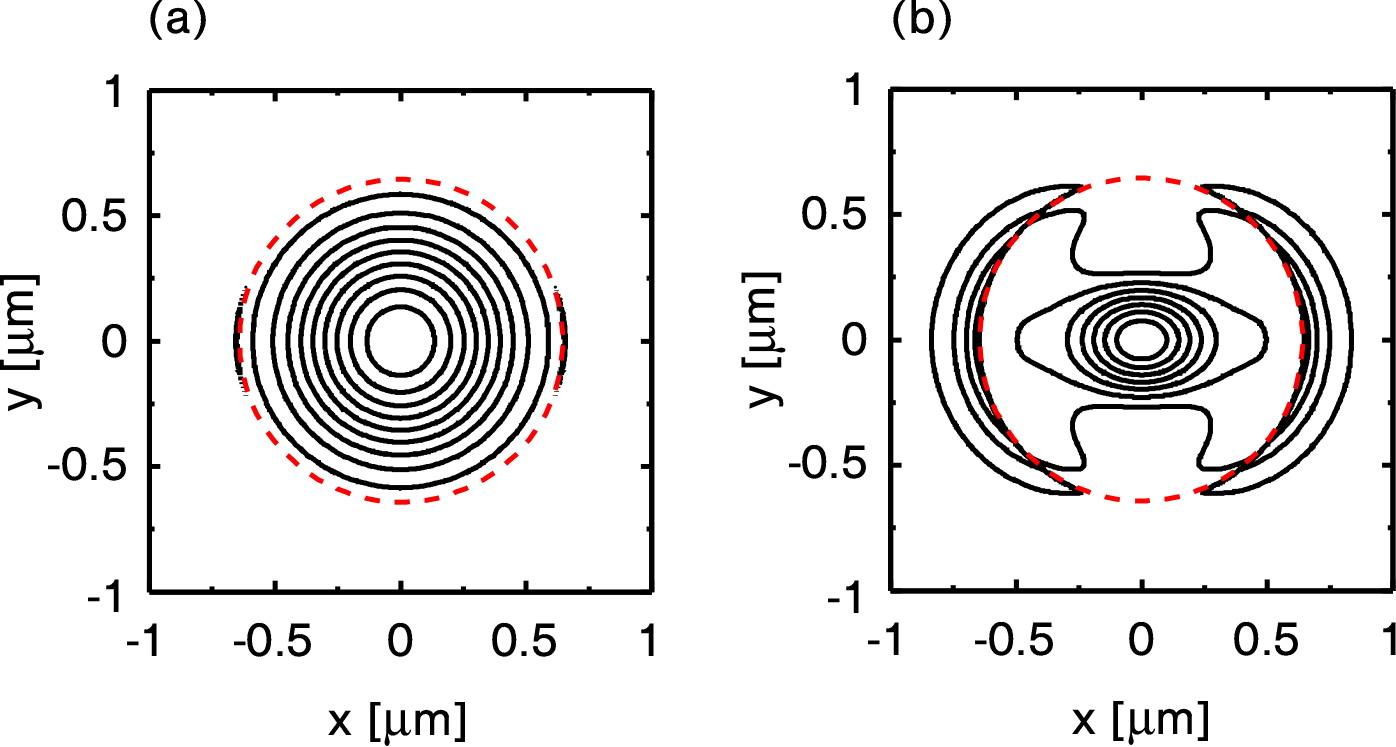}
}
    \caption{Contour plot of the transversal intensity distribution of the HE$_{11}$ and HE$_{12}$ modes. The dashed line indicates the nanofiber surface. The calculation was performed for a 1.3-$\mu$m diameter nanofiber and linearly polarized light in the $x$-direction at a wavelength of 850~nm.
}\label{fig:bonn_interferometer_modes}
\end{figure}

The amount of light outside the fiber body is different for the HE$_{11}$ and HE$_{12}$ modes. For example, for a waist diameter of 1.3\um and a wavelength of 850~nm, 2\percent of the optical power of the HE$_{11}$ mode and 31\percent of the optical power of the HE$_{12}$ mode is propagating in the evanescent field. Figure~\ref{fig:bonn_interferometer_modes} shows the calculated intensity distribution for both modes. As a result, a dispersive medium around the fiber will influence the phase accumulated by the HE$_{12}$ mode more strongly than that for the HE$_{11}$ mode and thus, create a measurable phase difference.

We have implemented and tested such an interferometer. For this purpose, we use a Fibercore SM800 fiber with a pure fused silica cladding, which is single-mode for wavelengths above 800~nm. The pulling rig was programmed to produce a taper slope of 16~mrad and a waist of 1.3\um diameter and 3~mm length. After pulling, the interferometer was tested in the pulling rig by coupling a single-frequency diode laser at 852~nm into the input pigtail of the fiber. The output was detected with a photodiode, and the stretcher stage of the pulling rig was used to elastically stretch the fiber by a few ten micrometers. 

\begin{figure}[h]
\centering
  \resizebox{0.48\textwidth}{!}{%
  \includegraphics{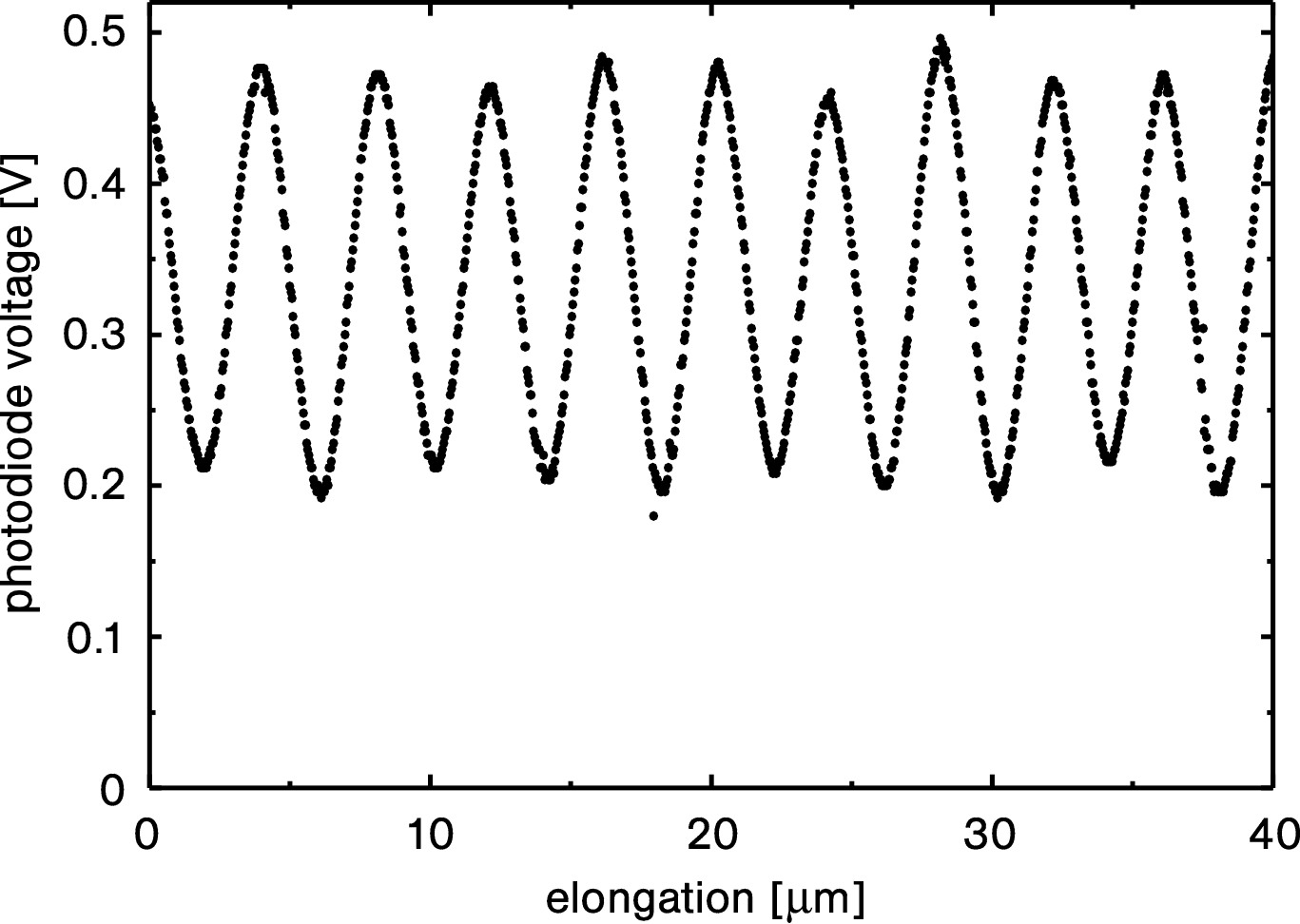}
}
    \caption{Interference fringes due to stretching of the fiber. The elongation of the waist leads to an additional phase shift between the HE$_{11}$ and the HE$_{12}$ modes. The slight modulation of the sinusoidal signal may indicate a weak presence of further modes possibly excited due to nanofiber asymmetries.
}\label{fig:Fig_int}
\end{figure}

Figure~\ref{fig:Fig_int} shows the resulting interference signal between the  HE$_{11}$ and the  HE$_{12}$ modes \cite{orucevic}. The changes in the phase difference are expected to be given by:

\begin{equation}
	\Delta (\varphi_{11} - \varphi_{12}) = (\beta_{11} - \beta_{12}) \Delta L + \Delta(\beta_{11} - \beta_{12}) L,
\end{equation}
where the term $\Delta(\beta_{11} - \beta_{12})$ takes into account the influence of both the Poisson and photoelastic effect on the propagation constants. The Poisson effect leads to the stretch-induced change of the fiber diameter. This, according to Fig.~\ref{fig:bonn_modes}, results in the change of the propagation constants of both of the propagating modes. The photoelastic effect explains the change of the material refractive index under strain. This change, in turn, is also influencing the propagation constants \cite{oft}.

The measurement outcome demonstrates the realization of a single-nanofiber-based interferometer with good interference contrast. A possible application of this device is to measure variations of the refractive index of the surrounding medium. The latter will differently change the propagation constants $\beta$ and thus lead to interference fringes which can then be traced.

% ====================== 5 Conclusions ==========================
\section{Outlook}
\label{sec:conclusions}

In this review, we have shown applications of fiber-based absorption and fluorescence spectroscopy. Our spectroscopic technique could be useful for bio-sensing, extending the possible applications from label-free detection by absorption measurements \cite{fan} to label-based detection using fluorescence measurements \cite{marazuela}. The fact that we are able to simultaneously record absorption and fluorescence spectra for a given surface coverage, in conjunction with the stable and reproducible fluorescence collection efficiency, makes our method ideal for analyzing and quantifying dynamic processes at or near the nanofiber surface. Excitation and fluorescence collection via a single fiber mode yields an entirely fiber-based method that can be used for spectroscopic studies at remote locations.   

We have also demonstrated our first results on the implementation of a nanofiber-based interferometer. The next goal is to perform dispersive detection on emitters adsorbed at the fiber waist and to compare the results with the absorption method. Further, the dispersive method should allow us to detect species that do not absorb in the available wavelength range or to carry out measurements on sensitive molecules while avoiding photobleaching.

\begin{acknowledgments} 
This work was supported by the Deutsche Forschungsgemeinschaft DFG (Research Unit 557), the Volkswagen Foundation (Lichtenberg Professorship), the European Science Foundation (European Young Investigator Award), and the European Comission (STREP CHIMONO).
\end{acknowledgments} 

% ====================== Bibliography ==========================

\end{document}